\documentclass[12pt]{article}
\usepackage{graphicx}
\usepackage{epsfig}
\newcommand{\rom}[1]{\mathrm{#1}}

\newcommand{\beq}{\begin{equation}}
\newcommand{\eeq}{\end{equation}}
\newcommand{\beqa}{\begin{eqnarray}}
\newcommand{\eeqa}{\end{eqnarray}}
\newcommand{\beqar}{\begin{eqnarray*}}
\newcommand{\eeqar}{\end{eqnarray*}}

\newcommand{\labell}[1]{\label{#1}} 
\newcommand{\reef}[1]{(\ref{#1})}

\newcommand{\eg}{{\it e.g.,}\ }
\newcommand{\ie}{{\it i.e.,}\ }

\begin{document}

\setlength{\unitlength}{1mm}

\thispagestyle{empty}
\rightline{\small hep-th/0310008 \hfill}
\vspace*{3cm}

\begin{center}
{\bf \Large Black Rings, Supertubes, and a Stringy Resolution of
Black Hole Non-Uniqueness}\\
\vspace*{2cm}

{\bf Henriette Elvang}\footnote{E-mail: {\tt
elvang@physics.ucsb.edu}}
{\bf and Roberto Emparan}\footnote{E-mail: {\tt emparan@ub.edu}}

\vspace*{0.2cm}

{\it $^1\,$Department of Physics}\\
{\it UCSB, Santa Barbara, CA 93106}\\[.5em]

{\it $^2\,$Departament de F{\'\i}sica Fonamental, and}\\
{\it C.E.R. en Astrof\'{\i}sica, F\'{\i}sica de Part\'{\i}cules i Cosmologia,}\\
{\it Universitat de Barcelona, Diagonal 647, E-08028 Barcelona, Spain}\\[.5em]

{\it $^2\,$Instituci\'o Catalana de Recerca i Estudis Avan\c cats (ICREA)}\\
[.5em]

\vspace{1.5cm} {\bf ABSTRACT}  
\end{center} 

\noindent 
In order to address the issues raised by the recent discovery of
non-uniqueness of black holes in five dimensions, we construct a
solution of string theory at low energies describing a five-dimensional
spinning black ring with three charges that can be interpreted as
D1-brane, D5-brane, and momentum charges. The solution possesses closed
timelike curves (CTCs) and other pathologies, whose origin we clarify.
These pathologies can be avoided by setting any one of the charges, \eg the
momentum, to zero. We argue that the D1-D5-charged black ring, lifted to
six dimensions, describes the thermal excitation of a supersymmetric
D1-D5 supertube, which is in the same U-duality class as the D0-F1
supertube. We explain how the stringy microscopic description of the
D1-D5 system distinguishes between a spherical black hole and a black
ring with the same asymptotic charges, and therefore provides a
(partial) resolution of the non-uniqueness of black holes in five
dimensions.

\vfill \setcounter{page}{0} \setcounter{footnote}{0}
\newpage

\setcounter{equation}{0}
\section{Introduction}
\labell{intro}

Rotating black holes typically exhibit much richer dynamics than their
static counterparts, especially in more than four dimensions. Static
black holes present relatively small qualitative differences as we
increase the number of dimensions beyond four, but the inclusion of
rotation brings in a larger diversity that depends on the dimensionality
of spacetime \cite{MP}. A recent surprise has been the discovery of a
rotating black ring solution in five-dimensional vacuum gravity
\cite{ER}. The horizon of the black ring has topology $S^2\times S^1$,
and its tension and self-gravitational attraction are precisely balanced
by the rotation of the ring. Perhaps more strikingly, when the mass and
spin are within a certain range of values, it is possible to find a
black hole and two black rings with the same spin and mass\footnote{We
follow \cite{ER} and often use the term ``spherical black hole", or even
simply ``black hole", to abbreviate the phrase ``topologically spherical
black hole", as opposed to the black ring. Technically speaking, the
black ring is also a black hole, but we hope to cause no confusion.}.
Hence, the simplicity of four-dimensional black holes implied by the
celebrated uniqueness theorems does not extend to five-dimensional
stationary solutions. It is quite possible that the non-uniqueness of
higher-dimensional black holes is indeed wider than indicated by the
existence of the black ring solution \cite{harv,EM}.

String theory has provided a remarkably successful description of black
holes, including an account of their entropy in microscopic terms
\cite{rev}, so it is natural to investigate in this context the
implications of these novel features of black holes. In particular we
can infer a striking consequence: to consistently account for these
new solutions, there must exist string states that so far have gone 
unnoticed in the microscopic analysis of certain configurations with
fixed conserved charges. 

Our current understanding of neutral black holes within string theory is
based on the ``correspondence principle" \cite{corr}. According to it, a
black hole is identified with a highly excited string state that is
obtained by adiabatically decreasing the gravitational (string)
coupling. When the black hole horizon shrinks to string-scale size,
stringy corrections to gravity become too large to be neglected. At this
point, the black hole can be matched to a string state with
(parametrically) the same value for the entropy. So one may ask what are
the string states that correspond to a black hole and to a black ring
with the same spin and mass, and how they differ. Conversely, if we take
a highly excited string and increase the coupling, at some point
gravitational collapse will occur. For certain values of the initial
mass and spin, the string will be confronted with the dilemma of what
object it collapses into: a black hole or a black ring. 

Unfortunately, the details of the string/black hole transition are still
too poorly understood to enable us to see how, or even whether, string
theory can distinguish between a neutral black hole and a black ring.
The problem is further compounded by the fact that there do exist black
holes and black rings with the same mass, spin, {\it and area}. From this
perspective, the black ring is perhaps an unexpected complication.

The map between black holes and string states becomes much more precise
if we consider charged BPS states (extremal black holes), which are
protected by supersymmetry as we vary the coupling, or even for systems
that are slightly excited above the BPS ground state (near-extremal
black holes)\footnote{More 
than on supersymmetry, the success of the stringy
description relies on an AdS$_3$ structure near the horizon. This is
also present in the solutions we construct.}. Therefore, in order to
investigate the issues discussed above, we construct in this paper a
family of charged black ring solutions that, near the BPS extremal
limit, can be identified with configurations of D-branes. 

When lifted to six dimensions, black rings become {\it black tubes}. The
BPS ground states of the tubes that appear in our study belong to a
class that has received recent attention. They can be referred to,
generically, as {\it supertubes}. Perhaps the best known example is a
configuration of D0-branes and fundamental strings (F1) bound to a
cylindrical D2-brane tube \cite{stube1}. Its supergravity description,
in different dimensions, was obtained in \cite{stube2}. From the
viewpoint of the worldvolume theory of the D2 brane, the dissolved D0
and F1 appear as crossed magnetic and electric fields. These produce a
Poynting vector tangent to the circular sections of the tube, and this
gives the system an angular momentum. Other supertubes are obtained from
this one via U-duality transformations. For instance, T-duality along
the direction of the tube gives a helical D-string, which carries linear
momentum as it coils around the tube axis \cite{dhelix}. Further
dualizations give a six-dimensional configuration where a D1-D5 bound
state rotates on a cylinder. The D1 and the D5 share the longitudinal
direction of the tube, with the other four directions of the D5-brane
wrapping a four-torus. In the map from the supertube with D0-F1 charges
to the one with D1-D5 charges, the circular section of the original D2
tube is mapped into a ring of Kaluza-Klein monopoles, and the
longitudinal direction of the tube, which is compactified, corresponds
to the $U(1)$ fiber of the KK-monopoles. So the D1 and D5 branes are
bound to a KK-monopole tube. A supergravity solution for a BPS D1-D5
supertube was first obtained in \cite{d1d5}, and has been generalized
and extensively studied in a series of papers \cite{M0,LM1,LM2,LM3,LMM}
(even if only the last among these refers to them as supertubes). 

The D1-D5 bound state is capable of carrying a momentum wave along the
D1-D5 intersection. We 
construct a three-charge black ring, which,
in addition to D1 and D5 charges, carries momentum charge (along the
sixth-dimensional direction of the tube). However, this extra charge
results in pathologies such as closed timelike curves (CTCs) at all
points. In fact we find that the boost that generates the momentum is
incompatible with the KK-monopole structure in the tube. So the way in
which these CTCs appear is different from other recently studied
instances of CTCs in rotating systems. Setting one of the charges, \eg
momentum, to zero, eliminates the pathologies, so in order to study the
microscopic string description we focus on D1-D5-charged black tubes.
The solution with equal D1 and D5 charges is dual to the charged black
ring obtained recently in \cite{HE} (which can be interpreted as
carrying equal F1 and momentum charges).

The D1-D5-charged black tubes can be regarded as thermally excited
supertubes. The non-extremal solutions have regular, non-degenerate
horizons of finite area and, as in the neutral case, there exist black
holes and black rings with the same values of the mass, spin, and
charges. By exploiting the detailed knowledge of the microscopics of the
D1-D5 system, we find that the stringy descriptions of black holes and
black rings are indeed different. This indicates how string theory can,
in this case, resolve the duplicity between such objects. We must add,
though, that we do not understand yet how string theory distinguishes
between the two black rings that have equal asymptotic charges.

The remainder of the paper is organized as follows: In Section
\ref{neutral} we introduce the neutral black ring, with some
improvements of the original description in \cite{ER}. The new
three-charge black rings are given in Section \ref{chargeup}, and its
physical parameters are computed in Section \ref{params}. In Section
\ref{ctcs} we study further the structure of the three-charge black
rings, and find pathologies, including CTCs, from combining a boost with a
KK-monopole. Thereafter we set the momentum charge to zero. Section
\ref{extremal} describes the extremal limit of the D1-D5-charged ring,
and the connection with supertubes. In Section \ref{microview} we
discuss the microscopic description of spinning black holes and rings
with D1-D5 charges, and show that they have a very different description
as states of string theory. Section \ref{discuss} contains the
conclusions and outlook of this work. Appendix \ref{app:solutions}
provides the intermediate steps leading from the neutral black ring to
the three-charge black ring, while appendix \ref{app:BMPV} shows how the
BMPV black hole \cite{bmpv,tseytlin} can be obtained as an extremal
limit of our solutions.

\setcounter{equation}{0}
\section{Neutral black ring}
\label{neutral}

We generate the new family of charged black rings by performing a series of
transformations on the five-dimensional neutral black ring of \cite{ER},
so we begin with an extended review of this solution. It is a vacuum
solution with metric
\beqa
  \nonumber
  ds^2 &=& -\frac{F(x)}{F(y)} \left(dt+
     R\sqrt{\lambda\nu} (1 + y) d\psi\right)^2  \\
  &&
   +\frac{R^2}{(x-y)^2}
   \left[ -F(x) \left( G(y) d\psi^2 +
   \frac{F(y)}{G(y)} dy^2 \right)
   + F(y)^2 \left( \frac{dx^2}{G(x)} 
   + \frac{G(x)}{F(x)}d\phi^2\right)\right] 
\labell{ring0}
\eeqa
with
\beq
  F(\xi) = 1 - \lambda\xi \, ,
\qquad  G(\xi) = (1 - \xi^2)(1-\nu \xi) \, .
\labell{fandg}\eeq
This form of the solution is slightly different from the original one
used in \cite{ER,HE}. First, we have made the minor notational change
$1/A=R$, since it is more sensible to have a radius scale for the ring
rather than the ``acceleration parameter" $A$ (which came from the C-metric
origin of this solution). Below we will see in what sense is $R$ a
radius of the ring. Second, we have adopted the form of the
cubic $G(\xi)$ proposed in \cite{hongteo}, which yields simple explicit
forms for its three roots,
\beq
\xi_2=-1\,,\quad \xi_3=+1\,,\quad \xi_4=\frac{1}{\nu}\,.
\labell{roots}
\eeq
Finally, having the roots of $G(\xi)$ in this form suggests also to
rename the root $\xi_1$ of $F(\xi)$ as $\xi_1=\lambda^{-1}$. The
solution then has dimensionless parameters $\lambda$ and $\nu$, and a
length scale $R$. 

The variables $x$ and $y$ take values in
\beq
-1\leq x\leq 1\,,\qquad  -\infty<y\leq-1\,,\quad \lambda^{-1}<y<\infty\,. 
\labell{xyrange}\eeq
As shown in \cite{ER}, in order to balance forces in the
ring one must identify
$\psi$ and $\phi$ with equal period
\beq
\Delta\phi=\Delta\psi=\frac{4\pi \sqrt{F(-1)}}{|G'(-1)|}=
\frac{2\pi\sqrt{1+\lambda}}{1+\nu}\,.
\labell{phsiperiod}
\eeq
This eliminates the conical singularities at the fixed-point sets $y=-1$
and $x=-1$ of the Killing vectors $\partial_\psi$ and $\partial_\phi$,
respectively. 
There is still the possibility of conical singularities at $x=+1$.
These can be avoided in two manners. Fixing
\beq
\lambda=\lambda_c\equiv \frac{2\nu}{1+\nu^2}\qquad {\rm (black\; ring)}
\labell{lambring}
\eeq
makes the circular orbits of $\partial_\phi$ close off smoothly also at
$x=+1$. Then $(x,\phi)$ parametrize a two-sphere, $\psi$ parametrizes a
circle, and the solution
describes a black ring. Alternatively, if we set
\beq
\lambda=1\qquad {\rm (black\; hole)}
\labell{lambhole}\eeq
then the orbits of $\partial_\phi$ do not close at $x=+1$. Then
$(x,\phi,\psi)$ parametrize an $S^3$ at constant $t,y$. The solution is
the same as the spherical black hole of \cite{MP} with a single rotation
parameter. Both for black holes and black rings, $|y|=\infty$ is an
ergosurface, $y=1/\nu$ is the event horizon, and the inner, spacelike
singularity is reached as $y\to\lambda^{-1}$ from above. 

The parameter $\nu$ varies in
\beq\labell{nurange}
0\leq\nu<1\,.
\eeq
As $\nu\to 0$ we recover a non-rotating black hole, or a very thin black
ring. At the opposite limit, $\nu\to 1$, both the black hole and the
black ring get flattened along the plane of rotation, and at $\nu=1$
result into the same solution with a naked ring singularity. 

We will often find it convenient to work with $\lambda$ as an
independent parameter, to be eventually fixed to the equilibrium value
$\lambda_c$. 
If we allow for values of $\lambda$ other than \reef{lambring} or
\reef{lambhole}, then whenever 
\beq \nu <\lambda <1
\labell{ringcond}\eeq
we find a black ring solution regular on and outside the horizon, except
for a conical singularity on the disk bounded by the inner rim of the
ring ($x=+1$). If $\nu <\lambda <\lambda_c$ then the ring is rotating
faster than the equilibrium value, and there is a conical deficit
balancing the excess centrifugal force. If instead $\lambda_c<\lambda<1$
then the rotation is too slow and a conical excess appears. If $\lambda
\leq \nu$ the horizon is replaced by a naked singularity. We shall refer
to the solution with $\lambda=\lambda_c$ as the equilibrium, or balanced
black ring. All these qualitative features will remain unchanged for the
non-extremal charged rings that we study below. 

The mass, spin, area, temperature and angular velocity at the horizon
for these solutions are given
by\footnote{The temperature here is defined as the surface gravity of
the horizon divided by $2\pi$. We prefer not to use the Euclidean
approach, because there is no analytic continuation of the black ring
that produces a non-singular, real Euclidean solution.}
\beq
M_0=\frac{3\pi R^2}{4G}\frac{\lambda(\lambda+1)}{\nu+1}\,,\qquad
J_0=\frac{\pi
R^3}{2G}\frac{\sqrt{\lambda\nu}(\lambda+1)^{5/2}}{(1+\nu)^2}\,,
\labell{mandj}\eeq
\beq
\mathcal{A}_0 = 
  8 \pi^2 R^3 \frac{\lambda^{1/2} (1+\lambda)(\lambda - \nu)^{3/2}}
       {(1+\nu)^2 (1-\nu)}\,,\quad
  T_0 = \frac{1}{4\pi R}\frac{1-\nu}
  {\lambda^{1/2} (\lambda - \nu)^{1/2} }
     \, ,
\labell{areatemp}\eeq
\beq
\Omega_0=\frac{1}{R}\sqrt{\frac{\nu}{\lambda(1+\lambda)}}\,.
\labell{omega0}\eeq
We use the subscript ${}_0$ for quantities that refer to the neutral
ring, as opposed to the charged rings below. Formulas \reef{mandj},
\reef{areatemp} and \reef{omega0} are valid
in general for $\lambda$ and $\nu$ in the ranges \reef{nurange},
\reef{ringcond}, as long as there is no conical defect at infinity, \ie
when \reef{phsiperiod} is satisfied.

Focusing on the
dimensionless quantity
\beq 
\frac{27\pi}{32 G}\frac{J_0^2}{M_0^3}=\left\{
\begin{array}{ll}
 \displaystyle{\frac{2\nu}{\nu+1}} &\quad \textrm{(black hole)}\\
{}& {} \\
\displaystyle{\frac{(1+\nu)^3}{8\nu}} &\quad 
\textrm{(balanced black ring)}
\end{array} \right.
\labell{jonm}\eeq
one easily sees that for black holes it grows monotonically from $0$ to
$1$, while for (equilibrium) black rings it is infinite at $\nu=0$,
decreases to a minimum value $27/32$ at $\nu=1/2$ \cite{HE}, and then
grows to $1$ at $\nu=1$. This implies that in the range
\beq
\frac{27}{32}\leq \frac{27\pi}{32 G}\frac{J_0^2}{M_0^3}<1
\labell{nonuniquej}\eeq
there exist one black hole and two black rings with the same
value of the spin for fixed mass. This regime of non-uniqueness occurs
when the parameter $\nu$ takes values in
\beq
\sqrt{5}-2\leq \nu <1 
\labell{nonuniquenu}
\eeq
for black
rings, and in
\beq
\frac{27}{37}\leq \nu <1 
\labell{nonuniquenubh}
\eeq 
for black holes. Bear in mind that when we speak about non-uniqueness
we always refer to equilibrium black rings.

There is a limit in which many features of the black ring become
particularly clear. Take $\lambda$ and $\nu$
as independent parameters, and let both of them approach
zero at the same rate (so $\nu/\lambda$ remains finite). Then we obtain
a ring with a large spin for a given mass. The ring is very thin
(`hula-hoop'-like), and locally it approaches the geometry of a boosted
black string. To recover this limit, we focus on a region near the
horizon, of size small compared to $R$, and scale
\beq
R\to\infty,\qquad \lambda,\nu\to 0
\labell{blowup1}\eeq
while keeping $\nu R$ and $\lambda R$ finite. Define new parameters
$r_0$, $\sigma$, and coordinates $r$, $\theta$, $w$ that remain finite in the
limit,
\beqa
&&\nu R=r_0 \sinh^2\sigma,\qquad \lambda R = r_0
\cosh^2\sigma,\nonumber\\
&&r=-R\frac{F(y)}{y},\quad \cos\theta =x,\quad w=R\psi,
\labell{blowup2}\eeqa
(so $|y|$ is large) to find that the black ring \reef{ring0}
goes over to
\beq
ds^2=-\bar f \left( dt -\frac{r_0
\sinh 2\sigma}{2r \bar f} dw\right)^2+ \frac{f}{\bar f}
dw^2+\frac{dr^2}{f}+r^2 d\Omega_2^2\,,
\eeq
with
\beq f= 1-\frac{r_0}{r}\,,\qquad\bar f= 1-\frac{r_0 \cosh^2\sigma}{r}\,.
\labell{fbarf}
\eeq
This a boosted black string, with boost parameter $\sigma$. When
$\lambda\to \nu$ the boost becomes light-like. For the equilibrium ring
we require instead $\lambda=\lambda_c\to 2\nu$, and the boost is
$\sinh|\sigma|=1$. This happens to be a rather special value. The ADM
stress-energy tensor (see \eg \cite{rob}) of the boosted black string is
\beqa\labell{setensor}
T_{tt}&=& \frac{r_0}{4G}(1+\cosh^2\sigma)\,,\nonumber\\
T_{tw}&=& \frac{r_0}{4G} \sinh\sigma\cosh\sigma\,,\\
T_{ww}&=& \frac{r_0}{4G} (\sinh^2\sigma -1)\,.\nonumber
\eeqa
Here $T_{tt}$ and $T_{tw}$ are the energy and momentum linear densities
of the black string, while $T_{ww}$ is its pressure density. The
limiting boost of the balanced black ring is $\sinh|\sigma|=1$, so its
limit is a pressureless black string. This absence of pressure reflects
the delicate balance of forces that the black ring represents.
Furthermore, one can check that for the balanced black ring the limiting
mass and
spin are such that
\beq\labell{densities}
\frac{M_0}{2\pi R}\to T_{tt}\,,\qquad \frac{J_0}{2\pi R^2}\to T_{tw}\,.
\eeq
Since $T_{tt}$ and $T_{tw}$ are densities of mass and momentum per unit
length, we see that $R$ must be interpreted as the circle radius of very
thin rings, in the sense that, in the limit, the coordinate $w$ is
identified as $w\sim w+2\pi R$. So $2\pi R$ is the proper length
measured at a large transverse distance from the string. This is not the
same as the proper length at the horizon, which is
$\Delta\psi\,\sqrt{g_{\psi\psi}}|_{y=1/\nu}\to 2\pi
R\sqrt{g_{ww}}|_{r=r_0}=2\pi R \cosh\sigma$, and is expanded by the
pressure of the boost. Consistently with this interpretation, the area
per unit length of a thin large ring is $\mathcal{A}_0/2\pi R\to 4\pi
r_0^2 \cosh\sigma$, \ie it becomes equal to the area per unit length of
the boosted black string. Also, in this limit the angular velocity
\reef{omega0} is $\Omega_0\to R^{-1}
\tanh\sigma$, so $\Omega_0 R$ becomes the boost velocity.

Eq.~\reef{densities} for the mass assumes that the ring is balanced, \ie
pressureless. For unbalanced rings the mass gets an additional
contribution from the pressure of the conical defect disk.

\setcounter{equation}{0}
\section{Charging up the ring}
\label{chargeup}

The method to build the three-charge black ring does not differ in
essence from the one that gave the three-charge rotating black hole in
five dimensions \cite{blmpsv,cv}. The final solution is given below in
eqs.~\reef{it3}-\reef{NS5ch}, for the case where the charges correspond to
fundamental strings (F1), NS5-branes and momentum $P$. Here and in
appendix \ref{app:solutions} we provide an outline and some of the
intermediate steps leading to it.

A rotating charged black ring with a single electric charge was
constructed in \cite{HE} as a solution to low energy heterotic string
theory. The charged solution was generated from the neutral
solution \reef{ring0} by application of a Hassan-Sen transformation
\cite{HS}. This involves a boost in an internal direction, and
the first step in generalizing the charged black ring of \cite{HE} is 
to realize that such a boost can be imitated by Lorentz boosts and
T-duality in an extra spatial direction followed by a Kaluza-Klein
reduction to five dimensions. We elaborate on this in the following as
we derive a solution for a black ring with two charges.  

Add to the metric \reef{ring0} a flat sixth dimension $z$. We also add
four flat directions $x^i$, $i=6,7,8,9$, wrapped on a $T^4$, but these
will play a more passive role.
A Lorentz boost,  
$dt \to \cosh{\alpha_5} \, dt + \sinh{\alpha_5} \, dz$ and
$dz \to \sinh{\alpha_5} \, dt + \cosh{\alpha_5} \, dz$,
gives the solution linear momentum in the $z$-direction, and
subsequent T-dualization of the $z$-direction exchanges the momentum
with a fundamental string charge from the 3-form flux \cite{HHS}. The
choice of subscripts in the boosts $\alpha_i$ will become clearer
later. We can then apply another Lorentz boost $\alpha_1$ in the
$z$-direction to get a solution with both charge and momentum. The
resulting six-dimensional black ring (or rather black tube) solution is
given in appendix \ref{app:solutions} in eqs.\
\reef{6d2ch}-\reef{Bpsiz}. It is a solution to the classical equations
of motion obtained from the action of the low energy NS-NS sector of the
superstring compactified on $T^4$,
\beq
  S_6 = \frac{1}{2 \kappa_6^2}
   \int d^6x \, \sqrt{-\hat{g}} \, e^{-2\Phi} 
   \left( R^{(6)} + 4 (\nabla \Phi)^2 - \frac{1}{12} \hat{H}^2\right)
   \, .
\eeq
Compactify the $z$-direction on a circle of radius $R_z$.
A Kaluza-Klein reduction along the $z$-direction, 
using the ansatz 
\beq
  \hat g_{MN}dx^M dx^N= g_ {\mu\nu}dx^\mu dx^\nu+
  e^{2\sigma}\big(dz + A^{(n)}_\mu dx^\mu\big)^2 \, ,
\eeq
where $x^M=(x^\mu, z)$ and $e^{2\sigma} = g_{zz}$, leads to a 
five-dimensional action 
\beqa
  \nonumber
  S_5 &=& \frac{1}{2 \kappa_5^2}
  \int d^5x \sqrt{-g} e^{-2\Phi+\sigma}\Bigg(
  R^{(5)} + 4 (\nabla\Phi)^2-4\nabla\Phi\nabla\sigma
  -\frac{1}{12} H^2 \\[2mm] 
  && \hspace{4.8cm}
  -\frac{1}{4}e^{2\sigma} \big(F^{(n)}\big)^2 
  - \frac{1}{4}e^{-2\sigma} \big(F^\rom{(1)}\big)^2 \Bigg) \, ,
  \labell{actionS5}
\eeqa
with $\kappa_5^2 =\kappa_6^2/(2 \pi R_z)$.
In \reef{actionS5},
$F^{(n)}_{\mu\nu} = 
\partial_\mu A^{(n)}_\nu -\partial_\nu A^{(n)}_\mu$, and we have
defined 
$A^\rom{(1)}_\mu = \hat{B}_{\mu z}$, so that 
$F^\rom{(1)}_{\mu\nu} = \partial_\mu A^\rom{(1)}_\nu 
-\partial_\nu A^\rom{(1)}_\mu
= \hat{H}_{z\mu\nu}$. 
The 3-form flux involves a Chern-Simons term and is given by
\beqa
  H_{\mu\nu\rho}
  &=& (\partial_\mu \hat{B}_{\nu\rho} 
  - A^{(n)}_\mu F^\rom{(1)}_{\nu\rho})
  + \rom{cyclic~permutations} \, .
  \labell{Hflux}
\eeqa
By setting $\sigma = 0$ and $F^\rom{(1)} = F^{(n)}$ one obtains a
consistent truncation of the theory \reef{actionS5} and the resulting
action is that of the heterotic string in five dimensions with only
one $U(1)$ subgroup included (see also \cite{DNP}). Defining the
effective dilaton $\Phi_\rom{eff} = \Phi - \sigma/2$,
the Einstein frame metric is
$g^\rom{E}_{\mu\nu} = e^{-\frac{4}{3}\Phi_\rom{eff}} g_{\mu\nu}$

The five-dimensional solution obtained by KK reducing the solution
\reef{6d2ch}-\reef{Bpsiz} describes a rotating black ring with two
charges, one from the KK gauge field $A^\rom{(n)}_\mu$ and the other
from the gauge field $A^\rom{(1)}_\mu$. Including in the metric
\reef{6d2ch} the four flat directions $x^i$, $i=6,7,8,9$, we can view
it as a solution with fundamental string charge $F1(z)$ (corresponding
to $\alpha_5$) and momentum in the $z$-direction $P(z)$ (from the
boost $\alpha_1$). 

For the special case where the boost parameters are $\alpha_5 = \alpha_1$,
the gauge fields become identical and $\sigma = 0$, so this is a
solution of the low energy heterotic string action. Up to
normalizations of the gauge fields, this is exactly the solution with
one charge found in \cite{HE}. 

We now proceed to find a solution with three charges. We continue from
the solution \reef{6d2ch}, plus the four flat directions $x^i$, viewed
as a Type IIB solution with $P$ and F1 charges. A sequence of dualities
maps this solution to a supergravity solution describing the
D1-D5-system. Boosting the resulting system in the direction $z$ before
KK reducing will then give us a black ring with three charges. We
describe the procedure briefly in this section and leave the details for
appendix \ref{app:solutions}. 

S-dualizing takes the black tube solution with
[$P(z)$, F1$(z)$] charges to a solution with charges
[$P(z)$, D1$(z)$]. T-dualizing the four directions $x^i$, 
then gives charges [$P(z)$, D5$(z6789)$], and S-dualizing again takes
us to a [$P(z)$, NS5$(z6789)$] system. Now T-dualizing $z$ converts the
momentum into F1 charge, giving a solution [F1$(z)$, NS5$(z6789)$] of
Type IIA. A T-dualization in any one of the $x^i$ direction is trivial
and takes us to Type IIB, and we can then S-dualize the solution to
get a tube solution with [D1$(z)$, D5$(z6789)$] charges.

For either of the solutions [F1$(z)$, NS5$(z6789)$] or [D1$(z)$,
D5$(z6789)$] we can apply a boost with parameter $\alpha_n$ in the
$z$-direction and then KK reduce to five dimensions. In appendix
\ref{app:solutions}, we give the metric and fields for the
[$P(z)$, F1$(z)$, NS5$(z6789)$] solution. Reducing \reef{sixd} to five
dimensions we find a black ring solution of \reef{actionS5} with
Einstein frame metric
\beqa  \labell{it3}
    \nonumber
    ds_5^2 &=& 
    - \frac{1}{(h_5 h_1 h_n)^{2/3}}\frac{F(x)}{F(y)} 
    \Bigg( dt  
    - \sqrt{\lambda\nu} R(1+y) 
    \cosh{\alpha_5}\cosh{\alpha_1}\cosh{\alpha_n}\, d\psi \\[2mm]
    && \hspace{5cm}
    + \sqrt{\lambda\nu} R(1+x) 
    \sinh{\alpha_5}\sinh{\alpha_1}\sinh{\alpha_n}\, d\phi \Bigg)^2
     \qquad \quad \\[2mm]
    \nonumber
    && \hspace{0.8cm}
    +(h_5 h_1  h_n)^{1/3}\frac{R^2}{(x-y)^2}
    \Bigg[ -F(x) \left( G(y) d\psi^2 +
    \frac{F(y)}{G(y)} dy^2 \right) \\[2mm]
    \nonumber && \hspace{5.2cm}
    + F(y)^2 \left( \frac{dx^2}{G(x)} 
    + \frac{G(x)}{F(x)}d\phi^2\right)\Bigg] \, ,\qquad
\eeqa
where we have defined
\beqa
  h_i(x,y) &\equiv&\nonumber
\frac{ F(y)\cosh^2\alpha_i- F(x)\sinh^2\alpha_i}{F(y)}\\
 &=& 1 + \frac{\lambda(x-y)}{F(y)} \sinh^2{\alpha_i} \, .
  \labell{hdelta}
\eeqa
The dilaton $\Phi$ and the extra scalar $\sigma$ are given by
\beq
  e^{-2\Phi} = \frac{h_1(x,y)}{h_5(x,y)} \, ,
  \hspace{1cm}
  e^{2\sigma} = \frac{h_n(x,y)}{h_1(x,y)} \, .
\eeq
The gauge fields are 
\beqa
    A^\rom{(1)}_t &=& \frac{(x-y) \lambda \sinh{2\alpha_1}}
                    {2 F(y) h_1(x,y)} \\[2mm] 
    A^\rom{(1)}_\psi &=&     
    \frac{\sqrt{\lambda\nu} R(1+y) F(x) 
          \cosh{\alpha_5}\sinh{\alpha_1}\cosh{\alpha_n}}
         {F(y) h_1(x,y)} \\[2mm]
    A^\rom{(1)}_\phi &=&     
    -\frac{\sqrt{\lambda\nu} R(1+x) 
          \sinh{\alpha_5}\cosh{\alpha_1}\sinh{\alpha_n}}
         {h_1(x,y)} \, ,
   \labell{A1}  
\eeqa
and 
\beq
  A^{(n)}_\mu = A^\rom{(1)}_\mu[\alpha_1 \leftrightarrow \alpha_n]
  \, . 
\eeq
There is an antisymmetric tensor field with components
\beqa
    B_{t\psi} &=&     
    -\frac{\sqrt{\lambda\nu}R(1+y) F(x) 
           \cosh{\alpha_5}\sinh{\alpha_1}\sinh{\alpha_n}}
         {F(y) h_1(x,y)} \\[2mm]
    B_{t\phi} &=&     
    \frac{\sqrt{\lambda\nu}R(1+x) 
           \sinh{\alpha_5}\cosh{\alpha_1}\cosh{\alpha_n}}
         {h_1(x,y)} \\[2mm]
    B_{\psi \phi} &=& 
    -\frac{1}{2}\lambda R^2 \sinh{2\alpha_5}
      \left( \frac{G(x)}{x-y} + k(x) 
       + \frac{\nu F(x)(1+x)(1+y) \sinh^2{\alpha_1}}
              {F(y)h_1(x,y)} \right) \, , \qquad 
    \labell{NS5ch}
\eeqa
where $k(x) = x(1+\nu) - \nu x^2 + \rom{const}$. In five dimensions,
the 3-form flux $H$ given in \reef{Hflux} is dual to a 2-form field
strength $F^{(5)} = e^{-2\Phi_\rom{eff}} \star H$. $F^{(5)}$ can be
obtained from the gauge potential  
\beq
  A^{(5)} =  A^\rom{(1)}_\mu[\alpha_1 \leftrightarrow \alpha_5] \, .
\eeq
Each boost is associated to one kind of charge,
\beqa
\alpha_5 &&\longrightarrow \mathrm{NS5}\,,\nonumber\\
\alpha_1 &&\longrightarrow \mathrm{F1}\,,\\
\alpha_n &&\longrightarrow P\,.\nonumber
\eeqa
Under S-duality, the NS5 and F1 transform into D5 and D1. Setting any
one of the charges to zero, we recover the two-charge black ring.
Originally, this was obtained by a reduction of the [$P(z)$,
F1$(z)$]-system, but the three charges [$P$, F1, NS5] can be permuted by
U-duality transformations of the ten-dimensional solution. The
Einstein-frame metric
\reef{it3} is invariant under such transformations, however, the radii
of compactification and the dilaton do change. If we regard the
[$P(z)$, F1$(z)$, NS5$(z6789)$]-ring as a solution to type IIA
supergravity, then it can be lifted to a solution of 11D supergravity,
with an additional coordinate $\zeta$, of the form [$P(z)$,
M2$(z\zeta)$, M5$(z6789)$]. Reduction along the coordinate $z$ yields
another type IIA solution, with D0, F1 and D4 charges, which is also
T-dual to the D1-$P$-D5 solution. 

The particular
case of the solution \reef{it3} where all charges are equal,
$\alpha_5=\alpha_1=\alpha_n$, yields a solution to minimal $N=2$
supergravity in five dimensions\footnote{This solution was obtained
previously by E.\ Teo, and then discarded due to the pathological CTCs
to be described in sec.\ \ref{ctcs} \cite{oldsoln}.}.

One remarkable feature of the new solution \reef{it3} is that, with the
inclusion of the third charge the black ring has acquired angular
momentum along a second independent axis, the direction $\phi$ of the
2-sphere. We will see, however, that this is more a problem than a
virtue.

\setcounter{equation}{0}
\section{Physical parameters}
\label{params}

Many of the qualitative properties of non-extremal black rings are the
same as for neutral black rings --- a main distinguishing
feature will be discussed in the next section. The parameters $\nu$
and $\lambda$, and the coordinates $x$ and $y$, vary in the same ranges
as in Sec.~\ref{neutral}, and $\phi$ and $\psi$ are identified with the
same periods \reef{phsiperiod}. Again, the event horizon is located at
$y=1/\nu$, and $y>1/\nu$ defines the ergoregion.

If we make the choice $\lambda = 1$ of the spherical black hole class  
of solutions (see \reef{lambhole}), we recover a particular case of
the three-charge rotating black hole in five dimensions of
\cite{blmpsv,cv}\footnote{The solutions in \cite{blmpsv,cv} have an
  additional parameter corresponding to the second angular momentum of
  the seed neutral Myers-Perry black hole.}. 
The change from the coordinates used in this paper (appropriate for
the ring) to the more conventional Boyer-Lindquist-type coordinates,
can be found in \cite{ER}. 
When all charges are nonzero this solution has a smooth inner horizon
at $y= 1/\lambda = 1$, but if either one of the charges is set to
zero, the inner horizon becomes a curvature singularity. 

When $\nu<\lambda<1$, and up to an important issue to be discussed in
next section, the horizon is topologically $S^2 \times S^1$ and the
solution is a spinning black ring, which is balanced and free of conical
singularities only when $\lambda=\lambda_c$ \reef{lambring}. For the
ring there is no smooth inner horizon; behind the event horizon at
$y=1/\nu$ the curvature blows up at $y=1/\lambda$.

The physical quantities for the charged black ring are computed
in the Einstein frame.
The ADM mass and angular momenta are
\beqa
  M &=& \frac{1}{3}M_0 \left(\cosh 2\alpha_5 
               + \cosh 2\alpha_1 
               + \cosh 2\alpha_n \right) \, ,\\[1mm]
  J_\psi &=& J_0 \cosh{\alpha_5} \cosh{\alpha_1} \cosh{\alpha_n} \, ,\\[1.5mm]
  J_\phi &=& -J_0 \sinh{\alpha_5} \sinh{\alpha_1} \sinh{\alpha_n} \, ,
\labell{mandj2}\eeqa
expressed in terms of the mass $M_0$ and the angular momentum $J_0$ of the
neutral solution \reef{mandj}. 

The ring couples electrically to the gauge fields and the three
corresponding charges are computed as
\beq
  Q_i  =  \frac{1}{2\Omega_3}
        \int_{S^3~\rom{at}~\infty} 
        \hspace{-7mm} e^{-2\Phi_i} \star F^{(i)}  \, ,
  \labell{genQ}
\eeq
where $\Omega_3=2\pi^2$ is the area of a unit three-sphere, $F^{(i)}$
is a field strength, $F^{(i)}=dA^{(i)}$, and 
$\Phi_1 = \Phi + \sigma/2$, $\Phi_5 = -\Phi_\rom{eff}$, 
and $\Phi_n = \Phi -3\sigma/2$, as can be seen from the action 
\reef{actionS5}. We find   
\beq
  Q_5 = \frac{4G}{3\pi} M_0 \sinh{2\alpha_5} \, ,
  \qquad
  Q_1  = \frac{4G}{3\pi} M_0 \sinh{2\alpha_1} \, ,
  \qquad
  Q_n  = \frac{4G}{3\pi} M_0 \sinh{2\alpha_n} \, .
  \labell{Qs}
\eeq
With this definition the six-dimensional momentum is
$P=\frac{\pi}{4G}Q_n$. Note that for all values of the parameters, the
charges and the mass
satisfy the inequality
\beq
  |Q_1| + |Q_5| + |Q_n| \le \frac{4G}{\pi}M \, ,
  \labell{QsandM}
\eeq
generalizing the bound found in \cite{HE}. 

It is clear that, again, there is a range of parameters where there
exist a black hole and two black rings with the same values of the mass,
spin, and the three charges $Q_i$. This happens for parameters in the
ranges \reef{nonuniquej}, \reef{nonuniquenu}, \reef{nonuniquenubh}.

The area of the horizon and its temperature admit also
simple expressions in terms of their values \reef{areatemp} 
for the neutral solution,

\beq
  \mathcal{A} = 
  \mathcal{A}_0
   \cosh{\alpha_5}\cosh{\alpha_1}\cosh\alpha_n \,,
\eeq
\beq
  T = \frac{T_0}
  { 
     \cosh{\alpha_5}\cosh{\alpha_1}\cosh\alpha_n} \, .
\eeq

The charges of the black ring are proportional to the total number of
branes and momentum quanta. To obtain these numbers, first we choose
units where $\alpha'=1$, and take for simplicity
the volume of the compact four dimensions $x^i$ to be $(2\pi)^4$, so the
five-dimensional Newton's
constant is
\beq\labell{fivedG}
G=\frac{\pi g_s^2}{4R_z}\,,
\eeq 
where $R_z$ is the radius of the direction $z$ along the
string/fivebrane intersection, and $g_s$ is the string coupling constant.
Since the solution is not extremal, there can be both branes and
antibranes, as well as left and right movers. The integer charges
measure the difference between their numbers.
For the D1-D5-$P$ black ring, the integer-quantized total numbers
of D1-branes, D5-branes, and momentum units are
\beq\labell{intd1d5n}
n_1-\bar n_1=g_s^{-1} Q_1,\quad n_5-\bar n_5=g_s^{-1} Q_5,\quad
n_L-n_R=n=R_z P=\frac{R_z^2}{g_s^2}Q_n\,.
\eeq
For the F1-NS5-$P$ ring the integer charges of F1 and NS5 are
$g_s^{-2}Q_1$, and $Q_5$, respectively.

The black ring also carries other local (dipole-type) charges. The full
structure of the solution will be uncovered in the next section.

In the same manner as we did for thin neutral rings, we can blow up the
region near the ring by taking the limits \reef{blowup1}, \reef{blowup2}.
This yields a charged black string in five dimensions, or a charged
black 2-brane if lifted to six dimensions. If we take the limit for the
F1-NS5-$P$ six-dimensional black tube \reef{sixd}, then the tube is
straightened out to a 2-brane spanning the coordinates $z,w$,
\beqa \labell{straightsixd}
    \nonumber
ds^2 &=& \frac{h_n}{h_1}
\Biggl( dz
+\frac{r_0\cosh^2\sigma\,\sinh 2\alpha_n}{2r h_n}\;dt-\frac{r_0\sinh 2\sigma
\cosh\alpha_5\cosh\alpha_1\sinh\alpha_n}{2rh_n}\; dw\nonumber\\
&&\hspace{32pt}+\frac{r_0\sinh 2\sigma
\sinh\alpha_5\sinh\alpha_1\cosh\alpha_n}{2h_n}(\cos\theta+1)\; d\phi
\Biggr)^2\nonumber\\
    &&- \frac{\bar f}{h_1 h_n}
    \Bigg( dt  
    +\frac{r_0\sinh 2\sigma}{2r\bar f} 
    \cosh{\alpha_5}\cosh{\alpha_1}\cosh{\alpha_n}\, dw \nonumber\\[2mm]
    && \hspace{58pt}
    + \frac{r_0\sinh 2\sigma}{2} 
    \sinh{\alpha_5}\sinh{\alpha_1}\sinh{\alpha_n}\;(\cos\theta+1)\, d\phi \Bigg)^2
     \qquad \quad \\[2mm]
    \nonumber
    &&
    +h_5 \left(\frac{f}{\bar f}
dw^2+\frac{dr^2}{f}+r^2 d\Omega_2^2\right)\,.
\eeqa
Here 
\beq
h_i=1+\frac{r_0}{r}\cosh^2\sigma\;\sinh^2\alpha_i\,.
\eeq 
Observe that we have left the boost parameters $\alpha_i$ fixed, so this
is not an extremal, nor near-extremal, limit. The charges $Q_i$ have
been simply scaled up, like the mass, as $R$, so as to keep a finite
charge per unit length of the ring\footnote{Close to extremality, this
is different than the near-horizon limit of the near-extremal black
ring. The details will be discussed elsewhere.}.

A calculation of the ADM stress-energy tensor as in \reef{setensor},
yields again vanishing pressure for the limit of the balanced ring. The
mass, momentum and charge densities agree, as in \reef{densities}, with
those for a string of (asymptotic) length $R$. The solution
\reef{straightsixd} can also be obtained by direct application of the
transformations of sec. \ref{chargeup} to the boosted black string.

\section{Closed Timelike Curves, and Kaluza-Klein-monopole tube}
\label{ctcs}

When all three charges are
non-zero the black ring metric \reef{it3} contains the one-form
\beq 
\mathcal{C}\equiv dt  - \sqrt{\lambda\nu} R(1+y) 
    \cosh{\alpha_5}\cosh{\alpha_1}\cosh{\alpha_n}\, d\psi
+ \sqrt{\lambda\nu} R(1+x) 
    \sinh{\alpha_5}\sinh{\alpha_1}\sinh{\alpha_n}\, d\phi
\eeq
with the effect that the orbits of $\partial_t$ are non-trivially
fibered over the $S^2\times S^1$ surfaces parametrized by
$(x,\psi,\phi)$. We should be careful to avoid Dirac-Misner string
singularities\footnote{A Dirac-Misner string is a gravitational analogue
of the Dirac string of gauge fields, first discussed by Misner in
\cite{misner} for the Taub-NUT solution.} in the geometry, which may
arise at the fixed-points of the orbits of $\partial_\psi$ and
$\partial_\phi$. Since $\mathcal{C}_\psi$ vanishes at the only possible
fixed-point set of $\partial_\psi$, $y=-1$, there is no singularity
there. However, if $(x,\phi)$ is to describe a two-sphere then
$\partial_\phi$ will have fixed-points at $x=\pm 1$. Now,
$\mathcal{C}_\phi=0$ at $x=-1$, but not at $x=+1$. We can remove the
Dirac-Misner singularity at $x=+1$ in a standard manner, taking a
different coordinate patch to cover this region. For some $\epsilon>0$,
take $t$ as the coordinate in
the patch $-1\leq x <\epsilon$, and $t'$ in $-\epsilon < x\leq 1$. In
the overlap
region $|x|<\epsilon$, relate them via
\beq
t'= t+ (2 \sqrt{\lambda\nu}\; R 
    \sinh{\alpha_5}\sinh{\alpha_1}\sinh{\alpha_n})\, \phi \,.
\labell{patchup}\eeq 
With this construction, the Dirac-Misner singularities are absent from
both poles $x=\pm 1$. But we have to pay a price: the matching
\reef{patchup} requires that $t$ (and $t'$) be identified with 
periodicity 
\beq\labell{tperiod}
\Delta t = (2 \sqrt{\lambda\nu}\; R 
    \sinh{\alpha_5}\sinh{\alpha_1}\sinh{\alpha_n})\, \Delta\phi\,
\eeq
(or an integer fraction of this). This introduces closed timelike curves
outside the horizon, even at infinity. So, either we have string
singularities, or we introduce naked CTCs: The solution with three
non-zero charges is pathological. This problem is instead absent from the
charged black hole solution, where $\partial_\phi$ has a fixed-point set
only at $x=-1$. 

To further clarify the origin of the problem, let us set the momentum
charge to zero, $\alpha_n=0$, so the pathology disappears, and consider
the solution lifted to six dimensions, as in \reef{sixd}. The relevant
structure of the solution is the same whether we consider the F1-NS5 or
D1-D5 black tubes. For later convenience we give the full ten-dimensional
D1-D5 solution,
with (string frame) metric
\beqa 
    \nonumber
ds^2 &=& \frac{1}{\sqrt{h_1 h_5}}
\Biggl( dz
-\sqrt{\lambda\nu}R(1+x)\sinh\alpha_5\sinh\alpha_1\; d\phi
\Biggr)^2\nonumber\\
    &&- \frac{1}{\sqrt{h_1 h_5}}\frac{F(x)}{F(y)} 
    \Bigg( dt  
    - \sqrt{\lambda\nu} R(1+y) 
    \cosh{\alpha_5}\cosh{\alpha_1}\, d\psi\Bigg)^2 \nonumber\\[2mm]
    \nonumber
    &&
    +\sqrt{h_1h_5}\frac{R^2}{(x-y)^2}
    \Bigg[ -F(x) \left( G(y) d\psi^2 +
    \frac{F(y)}{G(y)} dy^2 \right) 
    + F(y)^2 \left( \frac{dx^2}{G(x)} 
    + \frac{G(x)}{F(x)}d\phi^2\right)\Bigg]\nonumber\\
   &&+ \sqrt{\frac{h_1}{h_5}}\;\sum_{i=6}^9 dx^idx^i \, ,
\labell{d1d5}\eeqa
dilaton 
\beq
  e^{-2\Phi} = \frac{h_5(x,y)}{h_1(x,y)} \, ,
\eeq
and the nonzero components of the RR 2-form are
  \beqa
    C_{t\phi} &=&     
    -\frac{\sqrt{\lambda\nu}R(1+x) 
           \sinh{\alpha_5}\cosh{\alpha_1}}
          {h_1(x,y)} \\[2mm]
    C_{\psi \phi} &=& 
    \frac{1}{2} R^2 \lambda \sinh{2\alpha_5}
      \left( \frac{G(x)}{x-y} + k(x) 
       + \frac{\nu F(x)(1+x)(1+y) \sinh^2{\alpha_1}}
              {F(y)h_1(x,y)} \right) \qquad \\[2mm]
    C_{tz} &=& -\frac{(x-y) \lambda \sinh{2\alpha_1}}
                    {2 F(y) h_1(x,y)} \\[2mm] 
    C_{\psi z} &=&     
    -\frac{\sqrt{\lambda\nu} R(1+y) F(x) 
          \cosh{\alpha_5}\sinh{\alpha_1}}
         {F(y) h_1(x,y)} \, .
  \eeqa
Now we see that the direction $z$ is non-trivially fibered over the
$(x,\phi)$ two-sphere. This fibration is actually quite familiar: it is
a Hopf fibration due to the presence of a KK-monopole. The KK-monopole
extends along the $\psi$ direction (and also along the four compact $x^i$), so
we have a KK-monopole tube. In fact this is not unexpected. The D1-D5
black tube, can be transformed via
U-dualities into a D0-F1 black tube. As will become clearer later, the
extremal limit of all these solutions are supertubes. It is known that
the D0-branes and F-strings in a supertube configuration live in the
worldvolume of a cylindrical D2-brane. Mapping this D2 cylinder back
into the D1-D5 tube, one obtains a KK-monopole along
the $\psi$ circle with KK-direction $z$. This is the structure we have
identified directly above. The KK monopoles in the D1-D5
supertube have also been noted in \cite{LMM}. They are also present in
the non-extremal solutions, as evidenced by the fibration, even if a
horizon appears before the `nut' is reached. Note that the KK-monopoles
are absent when we consider spherical black hole solutions.

As before, removal of Dirac-Misner strings from the fibration requires
that we identify $z$ with period
\beq
\Delta z \equiv 2\pi R_z= \frac{2 \sqrt{\lambda\nu}\; R 
    \sinh{\alpha_5}\sinh{\alpha_1}}{n_\rom{KK}}\, \Delta\phi\,,
\labell{KKcond}\eeq
where $n_\rom{KK}$ is an integer that counts the number of KK-monopoles
that make up the tube (or the number of times the KK-monopole winds
around the tube). Since the KK-monopole is distributed on a circle, its
charge is not one of the conserved charges of the black ring. Instead,
it appears in the asymptotic region as a magnetic dipole of the field
$A^{(n)}$. The local charge can be measured by integrating the KK
magnetic flux across an $S^2$ that intersects the ring at a single
point, analogous to the calculation of the local D2 charge for the
supertube in \cite{stube2}.

It is actually the combination of the KK-monopole and the boost along
the KK-direction $z$ that is responsible for the pathological 
Dirac-Misner strings that we have found. To see this, consider the
solution for a KK-monopole in five dimensions,
\beq
ds^2=-dt^2+H^{-1}\left[dz +q(\cos\theta+1)\;d\phi\right]^2+H (dr^2+r^2
d\Omega_2^2)
\eeq
with $H=1+\frac{q}{r}$. The string at $\theta=0$ can be removed
with adequately chosen coordinate patches. Now boost it along the
direction $z$ with boost parameter $\alpha_n$, to find
\beqa\labell{boostedKK}
ds^2&=&\frac{H_n}{H}\left( dz-\frac{q\sinh
2\alpha_n}{2rH_n}\;dt+\frac{q\cosh\alpha_n}{H_n}(\cos\theta+1)\;d\phi\right)
^2\nonumber\\
&&-H_n^{-1}\left(dt-q\sinh\alpha_n(\cos\theta+1)\;d\phi\right)^2
+H(dr^2+r^2d\Omega_2^2)
\eeqa
where $H_n=1-\frac{q\sinh^2\alpha_n}{r}$. The essence of the fiber
structure of $z$ and $t$ over the $S^2$ in this solution can be
recognized in \reef{straightsixd}, and in general it is clear how the
Dirac-Misner strings at the poles of the $S^2$ force the periodicity of
the time $t$ in the same manner as we found for the three-charge black
ring. As a matter of fact, the KK-monopole fibration imposes
identifications in the geometry that are incompatible with the boost. So
besides the CTCs, the geometry \reef{boostedKK} does not even describe a
consistent fibration. This same problem is present for the three-charge
black tube.

The dual tube with D0, F1 and D4 charges (the D4
wraps the four internal directions $x^i$) exhibits the same
inconsistencies even if, as a solution to IIA supergravity, it does not
contain a KK-monopole. When lifted to eleven dimensions, this system is
described by M2-branes and M5-branes on a tube, carrying momentum along
a common intersection. Exactly the same solution is obtained by lifting
the type IIA ring with F1, NS5 and momentum charges. This implies that
the M-branes are again in the background of a KK-monopole tube, which is
incompatible with the momentum wave.

Note that these problems are of a different nature, and more 
serious,
than the CTCs found for the over-rotating charged black hole in five
dimensions, which appear as a result of the identifications that
compactify the solution from six to five dimensions \cite{ch1}. Undoing
these identifications and going to the universal covering then removes
the CTCs. In our case, the presence of the KK-monopole in the D1-D5 or
F1-NS5 black tube {\it forces} the direction $z$ to be compact ---
$n_\rom{KK}$ cannot be zero --- and also forbids the possibility of
boosting the solution.

So three-charge black rings appear to be unphysical. However, the
two-charge solutions obtained by setting any one of the charges to zero
are perfectly sensible: the D1-D5 solution does not have momentum, and
the D1-$P$ and D5-$P$ do not have a KK-monopole tube. The three-charge
solution allows us to obtain any of these two-charge solutions
immediately, and is at least a useful way of encoding all these U-dual
solutions.

\setcounter{equation}{0}
\section{Extremal limit}
\label{extremal}

An extremal limit of the solution is obtained by sending at least one
of the boost parameters $\alpha_i$ to $\infty$, while keeping the metric
and the corresponding charges $Q_i$ finite. From \reef{Qs} we see that
we must send $M_0\to 0$ while the products $M_0 e^{2\alpha_i}$ remain
finite. One possibility is to take $R\to 0$, with $\lambda$ and $\nu$
finite. We show in appendix \ref{app:BMPV} that in this way one always
recovers the extremal three-charge spherical black hole. 

Here we are more interested in the limit where $\lambda,\nu\to 0$ while
$R$ stays finite, which preserves the ring-like structure of the
solution. Since $\lambda\neq 1$, the possibility that this limit be
taken for spherical black holes is excluded. Also, it is possible to
have an extremal limit for the three-charge ring if $\lambda$ and $\nu$
go to zero at a different rate. But the extremal three-charge solution
that results presents the same pathologies that we found for generic
three-charge rings, and therefore we will not consider it further. In
the following we set the momentum charge to zero. It is easy to redo the
analysis when, instead, any of the other two charges is set to zero.

To obtain this extremal limit for the D1-D5 black tube, 
send $\alpha_1,\alpha_5\to\infty$ and
$\lambda,\nu\to 0$ while keeping
\beqa
\lambda e^{2\alpha_i} &=&\frac{2Q_i}{R^2}\,, \nonumber\\
\sqrt{\lambda\nu}e^{\alpha_5+\alpha_1}&=&\frac{8G}{\pi}\frac{J}{R^3}=
\frac{2g_s^2}{R_z}\frac{J}{R^3}\,,
\labell{limitqj}\eeqa
fixed. Note in the last line we have used \reef{fivedG}.

The extremal solution can be put in a convenient form with a change to
new coordinates $(x,\;y)\to
(r,\;\theta)$, 
\beq
r^2=R^2\frac{1-x}{x-y}\,,\qquad \cos^2\theta=\frac{1+x}{x-y}\,.
\eeq
Then the metric, written down as a ten-dimensional D1-D5
solution in string frame, is 
\beqa  \labell{extring}
    \nonumber
    ds^2 &=& 
-\frac{1}{\sqrt{h_1 h_5}}\left( dt  
+\frac{g_s^2}{R_z}\frac{J \sin^2\theta}{\Sigma}\,d\psi\right)^2    
+\frac{1}{\sqrt{h_1 h_5}} \left(dz 
- \frac{g_s^2}{R_z}\frac{J \cos^2\theta}{\Sigma}\,d\phi \right)^2
\\
&+&\sqrt{h_1 h_5}\left[\Sigma\left(\frac{d r^2}{r^2+R^2}+d\theta^2\right)+
(r^2+R^2)\sin^2\theta d\psi^2 + r^2\cos^2\theta d\phi^2\right]\\
\nonumber &+& \sqrt{\frac{h_1}{h_5}}\;\sum_{i=6}^9 dx^idx^i \, .
\eeqa 
with 
\beq
h_{i}=1+\frac{Q_i}{\Sigma}\,,\qquad
\Sigma\equiv  r^2+R^2 \cos^2\theta\,.
\labell{hsigbl}\eeq

When this extremal solution is obtained as a limit of balanced black
rings, $\lambda$ is not a free parameter, instead $\lambda\to 2\nu$ in
the limit, so from \reef{limitqj} we find that the spin
of the extremal ring is fixed to be
\beqa\labell{spinring1}
J^2&=&\frac{1}{2}\frac{R_z^2}{g_s^4}Q_1Q_5R^2\nonumber\\
&=&\frac{1}{2}\frac{R_z^2}{g_s^2}n_1n_5R^2\,,
\eeqa
where we use \reef{intd1d5n} and the fact that for the extremal solution
there are no anti-D1 or anti-D5 branes. As described in sec.~\ref{ctcs},
the D1 and D5 branes are rotating on top of a KK-monopole tube, which
imposes the periodicity condition \reef{KKcond} on the $z$ coordinate.
In the extremal limit \reef{limitqj}, equation \reef{KKcond} results in 
\beq\labell{KKcond2}
n_\rom{KK}R^2=\frac{g_s^2}{R_z^2}J\,.
\eeq
Together with \reef{intd1d5n}, this allows us
to write the equation \reef{spinring1} for the spin of the
extremal ring in terms of the numbers of strings, five-branes, and
KK-monopoles, as
\beq\labell{spinring2}
J= \frac{1}{2}\frac{n_1 n_5}{n_\rom{KK}}\,.
\eeq 

This two-charge extremal ring, or better, extremal tube in 6D, is in
fact a D1-D5 {\it supertube}, within the class of $1/4$-supersymmetric
solutions studied in \cite{LM3,LMM}. Another coordinate form, that
allows to
make contact with the (six-dimensional) supertube solutions in
\cite{stube2}, is obtained by changing $(r,\theta)\to (\rho_1,\rho_2)$,
\beq
\rho_1=\sqrt{r^2+R^2}\sin\theta\,,\qquad
\rho_2=r\cos\theta\,.
\eeq
Then $h_{1,5}$ take the same form as in \reef{hsigbl}, with
\beq
\Sigma =\sqrt{(\rho_1^2+\rho_2^2+R^2)^2-4R^2\rho_1^2}\,,
\eeq
and the metric is
\beqa  
    \nonumber
    ds^2 &=& 
    -\frac{1}{\sqrt{h_1 h_5}} 
    \left( dt  
+ \frac{4G}{\pi}\frac{2 J\rho_1^2}{\Sigma (\rho_2^2+R^2+\Sigma+\rho_1^2)}\, d\psi
\right)^2\nonumber\\
&+&\frac{1}{\sqrt{h_1 h_5}} 
    \left( dz  
- \frac{4G}{\pi}
\frac{J(\rho_2^2+R^2+\Sigma-\rho_1^2)}{\Sigma(\rho_2^2+R^2+\Sigma+\rho_1^2)}\, 
d\phi \right)^2\\
\nonumber &+&
\sqrt{h_1 h_5} 
\left(d\rho_1^2+\rho_1^2d\psi^2 + d\rho_2^2+\rho_2^2 d\phi^2\right)\\
\nonumber &+& \sqrt{\frac{h_1}{h_5}}\;\sum_{i=6}^9 dx^idx^i \, .
\eeqa 
Comparing to \cite{stube2}, we see that $h_{1,5}$ are harmonic
functions in the flat ${\bf R}^4$ of coordinates $(\rho_1, \psi, \rho_2,
\phi)$, with sources on a circle at $\rho_1=R$, $\rho_2=0$. These are then
D1$(z)$-D5$(z6789)$ branes distributed on a tube. 

In \cite{stube2} it was found that an upper bound on the spin of
supertubes follows from the requirement that $\partial_\psi$ does not
become timelike and hence CTCs do not appear. The place where this could
happen is close to the
tube, where the norm of $\partial_\psi$ becomes approximately, up to a
positive factor,\footnote{See \cite{stube2} for the expression including
all terms (in this reference, $g_s=1=R_z$).}
\beq
g_{\psi\psi}\propto -J^2+\frac{R_z^2}{g_s^4}Q_1 Q_5 R^2 \,.
\eeq
Then, the absence of CTCs implies an upper bound on the spin of the tube
\beq
J^2\leq \frac{R_z^2}{g_s^4}Q_1 Q_5 R^2 =\frac{R_z^2}{g_s^2}n_1 n_5 R^2\,,
\labell{spinbound}\eeq
or, using \reef{KKcond2}, 
\beq
J\leq \frac{n_1 n_5}{n_\rom{KK}}\,.
\labell{spinbound2}\eeq 
The extremal limit of the black ring solution satisfies \reef{spinring1}
and \reef{spinring2}, and hence does {\it not} saturate these bounds. As
in \cite{stube2}, we interpret the solutions that do not saturate the
spin bound as configurations where not all the D1 and D5 branes
contribute to the rotation. Some of them are simply superimposed on the
tube, together with those that contribute to the angular momentum. In
the extremal limit, there are no forces between supertubes and
(parallel) D1 or D5 branes, so the total energy of the configuration
depends only on the total number of D1 and D5 branes,
$M=\frac{\pi}{4G}(|Q_1|+|Q_5|)$.

If one fixes the conserved charges, or numbers of branes $n_1$ and $n_5$
in the solution, then the maximum possible spin is $J_\rom{max}=n_1
n_5$. Note that there are two different ways in which one can have
$J<J_\rom{max}$. The extremal D1-D5 solutions where
$J=n_1n_5/n_\rom{KK}$ with $n_\rom{KK}\geq 1$ were studied in
\cite{d1d5,LM1}, who showed that if $n_\rom{KK}=1$ the geometry near the
horizon is the non-singular global AdS$_3\times S^3$, while if
$n_\rom{KK}>1$ there is a conical deficit of the form (AdS$_3\times
S^3)/{\bf Z}_{n_\rom{KK}}$ (see also \cite{ME}). Alternatively, one may
have only a fraction of the branes contributing to the rotation, so the
bound \reef{spinbound2} is not saturated, as we have found for the
extremal limit of the black ring. The cases where $J<n_1n_5$ and
$n_\rom{KK}=1$ have been studied in \cite{LM2,LM3,LMM}. 
In general, the
solutions that do not saturate the spin bound \reef{spinbound2},
which include the extremal limit of the charged black ring, have a naked
strong curvature singularity at the ring. 

The D1-D5 supertubes with $J=n_1n_5/n_\rom{KK}$ were
constructed in \cite{d1d5} as a limit of the three-charge black hole. In
this case the solutions near extremality had null singularities instead
of horizons. In other words, this limit can not proceed through regular
near-extremal black rings that are continuously connected to the
supertube --- the gap for excitations is too large. We are interested in
approaching the supertubes through a sequence of
non-extremal solutions with regular horizons, so that there exist
near-extremal black rings arbitrarily close to the extremal, ground
state supertube. The near-extremal solutions are then viewed as
thermally excited states above the ground state. Let us first consider a
sequence of equilibrium black rings, which have regular horizons. As the
extremal limit is approached, we move toward supertubes with a
specific value of $J$, which is half the maximum spin for fixed $n_1$,
$n_5$ and $n_\rom{KK}$, \reef{spinring2}. But if we
allow for unbalanced black rings then we can obtain limiting supertubes
with 
\beq\labell{lessspin}
J=\frac{\nu}{\lambda}\frac{n_1n_5}{n_\rom{KK}}\,.
\eeq
Regularity of the non-extremal black ring horizon, up to a possible
conical singularity disk, requires $\nu<\lambda$. So we can find
limiting supertubes with any spin below the CTC-bound \reef{spinbound2}
from (generically unbalanced) near-extremal black rings. The particular
case where $\lambda\to\nu$ saturates the bound, although non-extremal
black rings with $\lambda=\nu$ have strong naked singularities. Extremal
tubes that exceed this bound are not connected to near-extremal black
rings. It is not clear to us why only the supertube with precisely half
the maximal spin can be excited to near-extremal black rings that are
balanced. We are not even sure that the conical singularity disk of
unbalanced rings can be given a meaning in the full string theory, but
in the extremal limit these conical singularities always disappear due
to a supersymmetric cancellation of forces.

\setcounter{equation}{0}
\section{Microscopic view: How strings tell a ball from a donut}
\label{microview}

For given D1 and D5 charges, there is a range of parameters
\reef{nonuniquej} in which there exist both black holes and black rings
that also have the same spin and mass. How can this duplicity of states
be understood within a microscopic stringy interpretation?

Let us first observe an important feature distinguishing D1-D5-charged
black holes and black rings --- recall that we are taking the momentum
charge to be zero. The spin of the non-extremal D1-D5 black
hole is bounded above by the energy above extremality. Near extremality
the bound is
\beq
J^2\leq \frac{\pi}{2G}(M-M_\rom{BPS})Q_1Q_5\,,
\eeq
with $M_\rom{BPS}=\frac{\pi}{4G}(|Q_1|+|Q_5|)$, and if this condition is
violated then the near-extremal solutions become naked singularities. So
the ground state of the D1-D5 spherical black holes (without momentum
charge) is not rotating. As we will
see, this reflects the fact that the angular momentum of the black hole
is carried by the same excitations that lift the system above the BPS
state. In contrast, 
we have seen that the spin of the black ring remains finite as the
energy above extremality decreases to zero. The ground state of the
spinning ring can have finite rotation. This indicates that the angular
momentum in this case is carried in a way different than 
in spherical black holes. 

In the following, we first identify the microscopic states that
correspond to the extremal limits of the solutions, and then we consider
their non-BPS excitations. As we have seen, the extremal BPS states are
generically supersymmetric D1-D5 
supertubes. Here we closely
follow the quite intuitive description of them given in \cite{M0}. For
simplicity we shall assume $n_\rom{KK}=1$, but the number of
KK-monopoles in the tube does not make much of a difference for our
discussion of black hole non-uniqueness. 

\begin{figure}[th]
\hskip1cm
\begin{center}
\epsfxsize=8cm\epsfbox{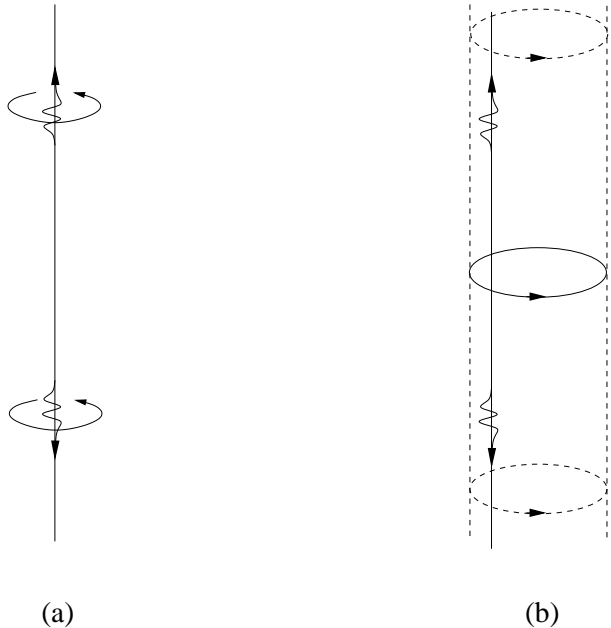}
\end{center}\hskip1cm
\caption{\small 
String theory interpretation of the spinning D1-D5 spherical
black hole (a) and black ring (b). The vertical direction corresponds to
the compactified sixth dimension, $z$, of the D1-D5 intersection. The
wavy excitations up and down it are left- and right-moving open strings
attached to the D1-D5 bound state. (a): In the spherical black hole, the
angular momentum is carried by the polarization of (equal numbers of)
left and right movers. When the left and right movers are turned off,
the spin disappears. (b): The D1-D5 black ring is a tubular
configuration where the angular momentum is due to the rotation around
the tube of strings of the D1-D5 system. The effect of the left and
right movers is only to excite the system above the BPS supertube ground
state, so they can be switched off while keeping the angular momentum
finite.}
\label{fig:d1d5}
\end{figure}

The 
dynamics of the bound state of $n_1$ D1-branes and $n_5$ D5-branes can
be effectively described in terms of a number $n_1 n_5$ of strings along
the common wrapped direction $z$ of the D1 and D5 branes. These strings
can be joined into a single string that is $n_1 n_5$ times longer. This
single long ``effective string'' can be in several ground states, one of
which has spin $1$ [\ie $(1/2,1/2)$ of the rotation group in five
dimensions $SU(2)\times SU(2)\simeq SO(4)$], a spin that is too small to
register as a macroscopically large angular momentum. But the string can
be excited by adding open strings that end on the D-branes and that
propagate along the D1-D5 intersection. These modes can carry a circular
polarization in the four non-compact directions and hence an angular
momentum of $SO(4)$. If there are a sufficiently large number of modes
with polarizations that add coherently, then the string will possess a
macroscopic rotation. When the open strings are all moving in the same
direction, left or right, then the system is still supersymmetric and we
obtain the microscopic description of the BMPV extremal rotating black
hole with D1, D5, and momentum charges \cite{bmpv}. However, we are
interested here in black holes without net linear momentum charge. In
this case we can still give the system an angular momentum by
exciting equal numbers of left and right movers, so the total linear
momentum $P=(n_L-n_R)/R_z$ is zero, and orienting their polarizations in the
same sense to give the system a macroscopic rotation. This configuration
is depicted in figure \ref{fig:d1d5}(a). 
Since we have excitations
moving in opposite directions along the string, the state is no longer
BPS. The supergravity description of this system is the non-extremal
rotating black hole with D1 and D5 charges and a spherical horizon of
finite area. It is the particular case of the solutions in
\cite{blmpsv,cv} where the momentum charge and the second
angular momentum are zero, and is also obtained from our solutions with
$\lambda=1$ and zero momentum $\alpha_n=0$. Close to extremality, its
Bekenstein-Hawking entropy can be precisely reproduced by a state
counting of the left and right moving modes \cite{blmpsv}. It is obvious
that, if we reduce to zero the number of left and right moving
excitations, then the rotation disappears. This agrees with the property
mentioned above that the spin of the D1-D5 spherical black hole
decreases to zero as we approach the extremal limit. 

There is, however, another way in which the D1-D5 state can acquire a
macroscopic rotation. As we said, a single string can, by itself, carry
only a unit of spin. But if a (macroscopic) fraction of the $n_1 n_5$
strings remain separate and do not join into a long string, then each of
them can contribute a unit of spin, which add up to give a macroscopic
angular momentum. These strings are still in a ground state, and they
form a rotating supertube configuration. When {\it all} the $n_1 n_5$
strings are separate and all of them spin, then one obtains the
supertube with maximal angular momentum $J_\rom{max}=n_1 n_5$. However,
we have seen that this is not the state that corresponds to the extremal
limit of the (equilibrium) black ring. Eq.~\reef{spinring2} (with
$n_\rom{KK}=1$) shows that in the latter only $n_1 n_5/2$ of the
effective strings contribute to the rotation, while the rest are
superimposed on the tube but do not add to the spin. This is then the
microscopic ground state of the D1-D5 black rings, where the branes 
themselves provide the rotation and the spin remains non-zero in the
extremal limit. 

We can also have
supertubes with any spin $J\leq J_\rom{max}$ and therefore a
different number of effective strings contributing the rotation, but, as
we saw in \reef{lessspin}, their associated near-extremal black rings
are not balanced and present conical singularity disks. The microscopic
explanation for this, and its consistency, are not yet clear to
us. We shall center our discussion around equilibrium black rings, for
which the issue of non-uniqueness is well-defined.

Having identified the microscopic string state for the extremal black
ring, now one can excite it above the BPS ground state. We do not yet
have the detailed nature of the thermal excitations of the ring, but it
seems plausible that, if we do not want to give the ring any net linear
momentum charge, then we can simply add equal numbers of left and right
moving open string excitations, see figure \ref{fig:d1d5}(b). The open
string modes are presumably carried by the effective strings that do not
contribute to the angular momentum, and which should join into a string
$n_1n_5/2$ times longer so as to lower the energy gap of excitations and
maximize the entropy. These excitations can be switched off smoothly
while the system retains its angular momentum, and this is indeed what
we observe in the supergravity solutions. 
We hope to clarify the
nature of these non-BPS excitations in subsequent work, but at present we
only need the fact that near-extremal black rings correspond, at weak
coupling, to thermally excited supertubes.

So string theory assigns different microscopic configurations to
spherical black holes and black rings. The identification of states that
we have performed is under good control if the systems remain close to
their supersymmetric ground state. This implies that the supergravity
solutions must be close to extremality, \ie $\nu$ close to zero, far
from the range of $\nu$ \reef{nonuniquenu} where the non-uniqueness of
the solutions is apparent. But it is natural to assume that, as we move
further away from the BPS state by adding more thermal excitations, the
corresponding supergravity solution changes continuously and preserves
its topology. Then, spherical black holes and black rings are still in
correspondence with thermally excited states of a single long D1-D5
string, and of supertubes, respectively. In this way, string theory can
resolve a black hole and a black ring with the same charges, spin and
mass. 

It remains unclear yet how string theory distinguishes between the two
black rings that also exist with the same asymptotic charges\footnote{We
may say that we are not able yet to distinguish a donut from a bagel.}.
As we move away from extremality, say, increasing the mass for fixed
spin and charges, the second black ring appears initially as a singular
solution, and always has less entropy than the first black ring (see
Fig.~3 in \cite{ER}). So the second solution seems to be a metastable
excitation of the system. In order to understand its complete
significance we may need a more detailed knowledge of the system far
from extremality than we currently have.

\setcounter{equation}{0}
\section{Discussion}
\label{discuss}

The addition of charge to the solutions of \cite{ER} has provided us
with a connection between two recently discovered types of rotating
objects in (super)gravity: black rings \cite{ER} and supertubes
\cite{stube2,d1d5,LM1}. Thus we have been able to recognize non-extremal
black rings as thermal excitations of supertubes, and this permits the
identification of black rings with states of string theory. An essential
point in this identification is that the different topologies of black
rings and black holes correspond to different topologies of the D-brane
configurations obtained at weak coupling. So string theory contains the
required states to account for at least some of the new supergravity
solutions, and this implies that the notion of black hole uniqueness is
not a necessary ingredient for the success in the microscopic account of
black hole entropy.

We have presented our discussion of the stringy description of black
holes and black rings in Sec.~\ref{microview} in the rather pictorial
terms of the ``effective string model", which is both convenient and
sufficient for identifying the origin of non-uniqueness. It is also
possible to employ the more precise language of the two-dimensional
conformal field theory that is dual to the D1-D5 system. The states of
the conformal theory that correspond to the extremal D1-D5 solutions
are, under spectral flow from the R to the NS sector, a particular class
of chiral primaries of the theory, and have been analyzed in detail in
\cite{LM2,LM3,LMM,ME}. The near-extremal black rings also possess a
near-horizon geometry that is asymptotically AdS$_3\times S^3$, and we
are currently investigating their dual CFT description. We hope to
report on this elsewhere.

A surprising feature of three-charge D1-D5-$P$ black rings is their
deadly pathological nature. We have seen that this is due to the
incompatibility of the KK-monopole structure, present whenever there are
both D1 and D5 branes rotating in a tube, and the boost that gives
momentum to the solution. 
An explanation of this issue from the AdS/CFT perspective should clarify
at the microscopic level why, if the thermal excitations in the
near-extremal black ring are momentum-carrying open strings, one can
have $n_L=n_R$ while $n_L\neq n_R$ seems to be more difficult or even
inconsistent.

We have illustrated this pathology with the example \reef{boostedKK} of
the combination of a KK-monopole and a boost in the KK-direction, but
this example does not imply that there is no consistent solution with
both electric and magnetic KK charges. Such solutions are actually known
\cite{emKK}, and they are free from the problems of \reef{boostedKK}.
There is also some understanding of them within string theory
\cite{shein}. However, it is unclear whether it is possible to add
momentum to the D1-D5 bound state in a KK-monopole tube in a similar
manner. It may also be interesting to see if a worldvolume analysis of
the dual system with D0-F1-D4 charges is able to shed further light on
this issue.

The concept of black hole uniqueness in four dimensions is well-defined
only for non-extremal solutions, with regular non-degenerate horizons.
The extremal limits of our charged black rings are $1/4$-supersymmetric
solutions whose horizons, even in the cases when they are non-singular,
are degenerate, and moreover have zero area. So even if a large number
of these solutions
do all have the same asymptotic charges\footnote{Classically, an
infinite number labeled by continuous parameters. Quantization should
discretize the spectrum.\label{ftn}}, they do not pose any problem to
the classical notions of uniqueness or no-hair. Recently, a uniqueness
theorem has been proven for supersymmetric black holes (with a
degenerate horizon) of minimal supergravity in five dimensions
\cite{harv}, but our two-charge solutions do not belong in this theory,
and furthermore they are typically singular. Nevertheless, they are
physically acceptable since they can be put in correspondence with
string states \cite{LM3,LMM}.

Questions like the interpretation of the non-uniqueness presented by the
duplicity of black rings remain open and beyond the control of the
near-extremal regime. Another similarly unresolved issue concerns the
lower bound on the spin of the ring. It seems that, if we add energy to
the ring while keeping its spin and charges fixed, there is a point at
which the ring cannot support itself against collapse any longer and
becomes unstable. Beyond this point there are no black ring solutions.
As shown in \cite{HE}, charge provides a repulsive force that allows for
equilibrium at smaller spins for a given mass, but there still remains a
lower bound on $J^2/M^3$. In fact, stability is an important open issue for
black rings (indeed, for all rotating black holes in more than four
dimensions \cite{EM}). Since thin black rings look locally like
(boosted) black strings, they are expected to suffer from the
Gregory-Laflamme instability \cite{GL}. The addition of charge, however,
may increase their stability, and perhaps near-extremal black rings
slightly excited above the BPS supertube ground state are stable.

While we can hardly expect that present techniques provide a detailed
understanding of non-uniqueness far from extremality, and in particular
for neutral black holes, we may still get some clues from what we have
learned so far. First of all, string theory does seem to possess the
additional states required to account for ring-like configurations.
Second, topology has been of crucial help in the matching to string
states: as the coupling is decreased, charged black holes and black
rings appear to go over into string configurations which themselves have
different topology, fig.~\ref{fig:d1d5}. The open string excitations
make the D1-D5 brane bound state somewhat fuzzier, and 
give a thickness
to the D1-D5 strings sketched in fig.~\ref{fig:d1d5}. As one moves
further away from extremality this fuzziness increases and becomes a
dominant feature. So this suggests a picture, presumably the simplest
one, where the topology of the solutions is preserved across the
correspondence point. Neutral rotating black holes should correspond to
highly excited (random-walk) strings that oscillate rapidly about a
center, and thus take the shape of a fuzzy string ball. If the spin of
these oscillations does not average to zero we get a net angular
momentum. Black rings, instead, would correspond to a less conventional
state where the excited string is wiggling about a circle and so has a
fuzzy donut-like shape. The angular momentum is carried by the coherent
orbital rotation of the string-donut around the circle, and it prevents
the state from collapsing into a string ball. We are not aware of any
previous discussion of such a string-donut state, but it should be
interesting to investigate whether it can exist, perhaps as a less
stable configuration than the string ball. String balls appear at weak
coupling as poles in string scattering amplitudes (see \eg \cite{DE}),
but it is not clear how to find string donuts as intermediate states.

Of course this is a very rough picture. It is far from obvious whether
it can explain issues such as the dimensionality dependence of these
effects, and the upper and lower bounds on the spin of black holes and
rings --- recall that the spin of fundamental string states is bounded
above $J\leq \alpha' M^2$ in any dimension. Moreover, topology cannot
resolve the duplicity among black rings. It will be interesting to
investigate whether we can at least obtain qualitative answers to these
questions.

We have considered configurations where the ``effective string'', \ie
the D1-D5 intersection, is transverse to the ring circle, thus forming a
tube. However, one may envisage a different situation where the
effective string is instead bent 
into a circle
to form itself the ring. In that case,
there would not be any net D1 and D5 charges, only dipole sources, and
the momentum that runs along the effective string would correspond to
the angular momentum. Such configurations, even when they are extremal,
are not expected to be supersymmetric. Rather than the D1-D5 bound
state, which would have to be smeared in one of the 
compact directions
transverse to a five-dimensional ring, it would be more interesting to
consider other systems that also yield effective strings and which are
more localized, such as the intersection of three M5-branes over a
string \cite{3m5}, or any of its U-duals. One such configuration was
discovered even before the neutral black ring was found. Ref.~\cite{tub}
described a solution of 11D supergravity where three M5-branes intersect
over a string loop, \ie a ring. The ring is not balanced, since it is a
static solution and there is no momentum running along the string loop.
Hence it has a conical singularity, but besides this, its near-horizon
geometry was shown to be AdS$_3$ times a distorted $S^2$.\footnote{When
the ring is balanced by immersing it in a fluxbrane, the conical
singularity and the distortion disappear and the geometry near the
horizon is exactly AdS$_3\times S^2$\cite{tub}.} The addition of a
certain amount of angular momentum should balance the ring and
presumably eliminate the distortion, and yield an extremal ring with a
regular horizon of finite area. Since this solution would not have any
asymptotic conserved charges other than mass and spin, it would provide
an infinite-fold violation of black hole uniqueness\footnote{See
footnote \ref{ftn}.}, a possibility also pointed out in \cite{harv}. So
it would be quite interesting to find it.

\section*{Acknowledgements} 
We would like to thank Gary Horowitz, Juan Maldacena, Harvey Reall, and
Edward Teo for useful discussions. 
HE would like the Niels Bohr Institute for hospitality during the final
stages of this work.
HE was supported by the Danish Research Agency and NSF grant
PHY-0070895. RE was supported in part by grants UPV00172.310-14497,
MCyT FPA2001-3598, DURSI 2001-SGR-00188, and HPRN-CT-2000-00131.  

\appendix

\setcounter{equation}{0}
\section{Towards the three-charge solution}
\label{app:solutions}

In this appendix we first give the solution for the black tube with
charges [$P(z)$,F1$(z)$]. Applying a sequence dualities, we then
obtain a black tube with charges [$P(z)$,F1$(z)$,NS5$(z6789)$], and
reduction of this to five dimensions yields the three-charge
spinning black ring of section \ref{chargeup}. 

The starting point is the ten-dimensional vacuum solution obtained by
adding to the black ring solution \reef{ring0} a flat direction $z$
as well as four directions $x^i$, $i=6,7,8,9$, on a torus $T^4$.
We apply a boost with parameter $\alpha_5$ in the $z$-direction,
T-dualize $z$, and then apply a second boost with parameter $\alpha_1$.
The resulting solution with $P(z)$ and F1$(z)$ charges associated with
$\alpha_1$ and $\alpha_5$, respectively, is described by the
string-frame metric
\beqa
  \nonumber
  ds_6^2 &=& -\frac{F(x) \hat h_1(x,y)}{F(y) h_5(x,y)} dt^2
    + \frac{(x-y)\lambda \sinh{2\alpha_1}}{F(y) h_5(x,y)} dt dz
    + \frac{h_1(x,y)}{h_5(x,y)} dz^2 \\[2mm]
  \nonumber
  && \hspace{0.6cm} 
    + \frac{2 \sqrt{\lambda\nu} R (1 + y) F(x) 
     \cosh{\alpha_5} \cosh{\alpha_1}}{F(y) h_5(x,y)} dt d\psi  \\[2mm]
  \labell{6d2ch}
  && \hspace{0.6cm} 
     + \frac{2 \sqrt{\lambda\nu} R (1 + y) F(x) 
     \cosh{\alpha_5} \sinh{\alpha_1}}{F(y) h_5(x,y)} dz d\psi   \\[2mm]
  \nonumber
  && \hspace{0.6cm}
    - \frac{F(x) \lambda \nu (y+1)^2 R^2 \cosh^2{\alpha_5}}
           {F(y) h_5(x,y)} d\psi^2 \\[2mm]
  && \nonumber \hspace{0.6cm}
   +\frac{R^2}{(x-y)^2}
   \left[ -F(x) \left( G(y) d\psi^2 +
   \frac{F(y)}{G(y)} dy^2 \right)
   + F(y)^2 \left( \frac{dx^2}{G(x)} 
   + \frac{G(x)}{F(x)}d\phi^2\right)\right] \, .
\eeqa
where $h_i(x,y)$ was defined in \reef{hdelta}, and $\hat
h_i(x,y)\equiv h_i(y,x)$. The full ten-dimensional solution also
includes four flat directions, $g_{ij} = \delta_{ij}$ for
$i=6,7,8,9$.

The dilaton is given by
\beq
  e^{-2\Phi} = h_5(x,y) \, ,
\eeq
and the antisymmetric tensor field has the following nonzero
components:
\beqa
  B_{t\psi} &=& 
   - \frac{\sqrt{\lambda\nu} R F(x) (1+y) \sinh{\alpha_5} \sinh{\alpha_1}}
          {F(y) h_5(x,y)} \, , 
  \labell{Btpsi} \\[2mm]
  B_{tz} &=& \frac{(x-y) \lambda \sinh{2\alpha_5}}
                          {2 F(y) h_5(x,y)} 
  \labell{Btz}\\[2mm]
  B_{\psi z} &=& 
   \frac{\sqrt{\lambda\nu} R F(x)(1+y) \cosh{\alpha_1}\sinh{\alpha_5}}
        {F(y) h_5(x,y)} 
  \labell{Bpsiz}\, .
\eeqa

Applying to the [$P(z)$, F1$(z)$] solution the sequence of dualities
\beq
  \rom{S}~+~\rom{T(6789)}~+~ \rom{S}~+~\rom{T(z)}
  ~+~\rom{boost}~\alpha_n
\eeq
we obtain the [$P(z)$,F1$(z)$,NS5$(z6789)$] solution. 
For each step, we outline the result of the dualities and list the
  types of charges:
\subsubsection*{\underline{S-dualize I:} $\to$ [$P(z)$, D1$(z)$]} 
  \begin{description}
  \item{Metric:} $ds'^2 = h_5^{1/2} (ds_6^2 + dx^i dx^i)$, with 
  $ds_6^2$ given in \reef{6d2ch}. 
  \item{Dilaton:} $e^{-2\Phi'} = h_5^{-1}$.
  \item{Forms:} ${C'}^{(2)}_{\mu\nu} = -B_{\mu\nu}$, with $B_{\mu\nu}$
  given in \reef{Btpsi}, \reef{Btz}, and \reef{Bpsiz}.
  \end{description}  
\subsubsection*{\underline{T(6789):} $\to$ [$P(z)$, D5$(z6789)$]}
  \begin{description}
  \item{Metric:} 
  $ds''^2 = h_5^{1/2} ds_6^2 + h_5^{-1/2}dx^i dx^i$.
  \item{Dilaton:} $e^{-2\Phi''} = h_5$.
  \item{Forms:} 
  ${C''}^{(6)}_{\mu\nu6789} 
  ={C'}^{(2)}_{\mu\nu}
  =-B_{\mu\nu}$.
  For the next step we need instead of ${C''}^{(6)}$ the Hodge dual
  2-form potential, $d{C''}^{(2)}=\star d{C''}^{(6)}$. We find
  \beqa
    {C''}^{(2)}_{t\phi} &=& (1+x) R \sqrt{\lambda\nu}
     \sinh{\alpha_5} \cosh{\alpha_1} \, ,
    \labell{Ctphi}\\
    {C''}^{(2)}_{\phi z} &=& - (1+x) R \sqrt{\lambda\nu}
     \sinh{\alpha_5} \sinh{\alpha_1}  \, ,
    \labell{Cphiz}\\
    {C''}^{(2)}_{\psi \phi} &=& -\frac{1}{2} R^2 \lambda \sinh{2\alpha_5}
      \left( \frac{G(x)}{x-y} + k(x) \right) \, ,
    \labell{Cpsiphi}
  \eeqa
  where $k(x) = x(1+\nu) - \nu x^2 + \rom{const}\,.$
  We have fixed the constants of integration so that
  ${C''}^{(2)}_{t\phi}$ and ${C''}^{(2)}_{\phi z}$ go to zero at infinity.
  \end{description}   
\subsubsection*{\underline{S-dualize II:} $\to$ [$P(z)$, NS5$(z6789)$]}
  \begin{description}
  \item{Metric:} $ds'^2 = h_5 ds_6^2 + dx^i dx^i$.
  \item{Dilaton:} $e^{-2\Phi'''} = h_5^{-1}$.
  \item{Forms:} $B'''_{\mu\nu} = {C''}^{(2)}_{\mu\nu}$, 
  with ${C''}^{(2)}_{\mu\nu}$
  given in \reef{Ctphi}, \reef{Cphiz}, and \reef{Cpsiphi}.
  \end{description} 
\subsubsection*{\underline{T(z) and boost $\alpha_n$:} 
  $\to$ [$P(z)$, F1$(z)$, NS5$(z6789)$]}
  \begin{description}
  \item{Metric:} 
    In the full ten-dimensional solution, there are four flat
    directions compactified on $T^4$: $g_{ij} = \delta_{ij}$  for
    $i,j = 6,7,8,9$. We can write the six-dimensional metric
    in a form appropriate for KK reduction along the $z$ coordinate
    (string frame):
    \beqa \labell{sixd}
      \nonumber
      ds^2 &=& \frac{h_n(x,y)}{h_1(x,y)}
      \Biggl( dz
      +\frac{(x-y)\lambda\sinh 2\alpha_n}{2F(y)h_n(x,y)}\;dt\nonumber\\
      &&\hspace{64pt}+\sqrt{\lambda\nu}R(1+y)\frac{F(x)}{F(y)
        h_n(x,y)}\cosh\alpha_5\cosh\alpha_1\sinh\alpha_n\; d\psi\nonumber\\
      &&\hspace{64pt}
-\sqrt{\lambda\nu}R(1+x)\frac{1}{
        h_n(x,y)}\sinh\alpha_5\sinh\alpha_1\cosh\alpha_n\; d\phi
      \Biggr)^2\\
      &&- \frac{1}{h_1(x,y)h_n(x,y)}\frac{F(x)}{F(y)} 
      \Bigg( dt  
      - \sqrt{\lambda\nu} R(1+y) 
      \cosh{\alpha_5}\cosh{\alpha_1}\cosh{\alpha_n}\, d\psi \nonumber\\[2mm]
      && \hspace{5cm}
      + \sqrt{\lambda\nu} R(1+x) 
      \sinh{\alpha_5}\sinh{\alpha_1}\sinh{\alpha_n}\, d\phi \Bigg)^2
      \qquad \quad \nonumber\\[2mm]
      \nonumber
      &&
      +h_5(x,y)\frac{R^2}{(x-y)^2}
      \Bigg[ -F(x) \left( G(y) d\psi^2 +
        \frac{F(y)}{G(y)} dy^2 \right) 
      + F(y)^2 \left( \frac{dx^2}{G(x)} 
        + \frac{G(x)}{F(x)}d\phi^2\right)\Bigg] \, . 
    \eeqa
  \item{Dilaton:} 
  \beq
  e^{-2\tilde{\Phi}} = \frac{h_1(x,y)}{h_5(x,y)} \, .
  \hspace{9.8cm}
  \eeq
  \item{Forms:} 
  \beqa
    \tilde{B}_{t\psi} &=&     
    -\frac{\sqrt{\lambda\nu}R(1+y) F(x) 
           \cosh{\alpha_5}\sinh{\alpha_1}\sinh{\alpha_n}}
          {F(y) h_1(x,y)} \\[2mm]
    \tilde{B}_{t\phi} &=&     
    \frac{\sqrt{\lambda\nu}R(1+x) 
           \sinh{\alpha_5}\cosh{\alpha_1}\cosh{\alpha_n}}
          {h_1(x,y)} \\[2mm]
    \tilde{B}_{\psi \phi} &=& 
    -\frac{1}{2} R^2 \lambda \sinh{2\alpha_5}
      \left( \frac{G(x)}{x-y} + k(x) 
       + \frac{\nu F(x)(1+x)(1+y) \sinh^2{\alpha_1}}
              {F(y)h_1(x,y)} \right) \qquad \\[2mm]
    \tilde{B}_{tz} &=& \frac{(x-y) \lambda \sinh{2\alpha_1}}
                    {2 F(y) h_1(x,y)} \\[2mm] 
    \tilde{B}_{\psi z} &=&     
    \frac{\sqrt{\lambda\nu} R(1+y) F(x) 
          \cosh{\alpha_5}\sinh{\alpha_1}\cosh{\alpha_n}}
         {F(y) h_1(x,y)} \\[2mm]
    \tilde{B}_{\phi z} &=&     
    -\frac{\sqrt{\lambda\nu}R(1+x) 
          \sinh{\alpha_5}\cosh{\alpha_1}\sinh{\alpha_n}}
         {h_1(x,y)} \, .
  \labell{BBphiz}
  \eeqa
  \end{description}   
The solution with D1, D5, and $P$
charges, in string frame, is obtained by multiplying \reef{sixd} by
$\sqrt{h_1(x,y)/h_5(x,y)}$, inverting the dilaton $\tilde\Phi\to
-\tilde\Phi$ and S-dualizing the NS-NS two-form to an RR two-form. The
three boosts $\alpha_5$, $\alpha_1$, $\alpha_n$ are each associated to
the D5, D1, and P charges, respectively.

\setcounter{equation}{0}
\section{BMPV black hole as an extremal limit}
\label{app:BMPV}

In section \ref{extremal} we mentioned a way to take the extremal limit
which involves sending $R\to 0$, while keeping $M_0 e^{2\alpha_i}$, \ie
$Re^{\alpha_i}$, finite. Since $R$ sets the scale for the metric, we
have to take the limit in such a way that we blow up (a part of) the
geometry to finite size. The change of coordinates
\beq
x=-1+\frac{2(1+\lambda)^2}{1+\nu}\frac{R^2\cos^2\theta}{r^2}\,,\qquad 
y=-1-\frac{2(1+\lambda)^2}{1+\nu}\frac{R^2\sin^2\theta}{r^2}\,,
\eeq
does precisely this: when $R\to 0$ keeping $r$ and $\theta$ finite, the
region near $x,y\simeq -1$ is blown up. The topology of the final
solution need not be the same as the original one, and in fact the
global problems that arise for black rings due to the fixed-point set of
$\partial_\phi$ at $x=+1$ disappear since $x=+1$ is pushed
away to infinity. 
Hence this limit can be considered even for three-charge black rings,
which as we saw in sec.~\ref{ctcs} are otherwise pathological.

The ten-dimensional extremal D1-D5-$P$ solution that one obtains is
\beqa  \labell{extbh}
    \nonumber
    ds^2 &=& 
    \frac{h_n}{\sqrt{h_1 h_5}} \left(dz +\frac{Q_n}{h_n r^2}dt
- \frac{g_s^2}{R_z}\frac{J \sin^2\theta}{h_n r^2}\,d\psi
-\frac{g_s^2}{R_z}\frac{J \cos^2\theta}{h_n r^2}\, d\phi \right)^2\\
 \nonumber   &&-\frac{1}{h_n\sqrt{h_1 h_5}}\left( dt  
+\frac{g_s^2}{R_z}\frac{J \sin^2\theta}{r^2}\,d\psi
+\frac{g_s^2}{R_z}\frac{J \cos^2\theta}{r^2}\,
d\phi \right)^2\\
&+&\sqrt{h_1 h_5}\left[d r^2+r^2 \left(d\theta^2+
\sin^2\theta d\psi^2 + \cos^2\theta d\phi^2\right)\right]\\
\nonumber &+& \sqrt{\frac{h_1}{h_5}}\;\sum_{i=6}^9 dx^idx^i \, ,
\eeqa 
with 
$\psi$ and $\phi$ rescaled to have periodicities $2\pi$, and
\beq
h_i=1 +\frac{Q_i}{r^2}\,.
\eeq
This is the extremal D1-D5-$P$ spherical rotating black hole, which upon
reduction yields the five-dimensional BMPV black hole
\cite{bmpv,tseytlin}. Note that we obtain this solution regardless of
the values of $\lambda$ and $\nu$, \ie of whether we are taking the
limit of a sequence of black holes or black rings. However, the limits
of black holes and black rings result into different ranges of values of the
final parameters. If all three charges are preserved in the limit, then,
we obtain the spin as
\beq
J^2=\left( \frac{27\pi}{32 G}\frac{J_0^2}{M_0^3}\right) n_1 n_5 n_L\,.
\labell{jbmpv}
\eeq
Along a sequence of black holes we get spins in the range $0\leq J^2\leq
n_1 n_5 n_L$, for which there are no CTCs outside the black hole
horizon. If instead we take an equilibrium black ring sequence, with
$\lambda=\lambda_c$, then for $0<\nu \leq \sqrt{5}-2$ we get spins $J^2
\geq n_1 n_5 n_L$, which is the over-rotating regime characterized by the
appearance of CTCs outside the horizon. As extremality is approached,
these CTCs do not have their origin in the periodic identification of
time \reef{tperiod}, but rather on $\psi$ becoming timelike outside the
horizon. It is also clear from
\reef{jbmpv} that there is a range of parameters for which a black hole
and two black rings limit to the same (under-rotating)
black hole at extremality.



\begin{thebibliography}{99}

\bibitem 
{MP} R.C.~Myers and M.J.~Perry,
Annals Phys.\  {\bf 172} (1986) 304.

\bibitem 
{ER} R.~Emparan and H.~S.~Reall,
Phys.\ Rev.\ Lett.\  {\bf 88} (2002) 101101
[arXiv:hep-th/0110260].

\bibitem 
{harv} H.~S.~Reall,
Phys.\ Rev.\ D {\bf 68} (2003) 024024
[arXiv:hep-th/0211290].


\bibitem 
{EM} R.~Emparan and R.~C.~Myers,
JHEP {\bf 0309} (2003) 025
[arXiv:hep-th/0308056].

\bibitem 
{rev} 
For a review, see, \eg A.W.~Peet,
arXiv:hep-th/0008241.

\bibitem 
{corr} 
L.~Susskind,
arXiv:hep-th/9309145.\\
G.~T.~Horowitz and J.~Polchinski,
Phys.\ Rev.\ D {\bf 55} (1997) 6189
[arXiv:hep-th/9612146].

\bibitem 
{stube1} D.~Mateos and P.~K.~Townsend,
Phys.\ Rev.\ Lett.\  {\bf 87} (2001) 011602
[arXiv:hep-th/0103030].

\bibitem 
{stube2} R.~Emparan, D.~Mateos and P.~K.~Townsend,
JHEP {\bf 0107} (2001) 011
[arXiv:hep-th/0106012].

\bibitem 
{dhelix}
J.~H.~Cho and P.~Oh,
Phys.\ Rev.\ D {\bf 64} (2001) 106010
[arXiv:hep-th/0105095].

\bibitem 
{d1d5} V.~Balasubramanian, J.~de Boer, E.~Keski-Vakkuri and S.~F.~Ross,
Phys.\ Rev.\ D {\bf 64} (2001) 064011
[arXiv:hep-th/0011217];\\
J.~Maldacena and L.~Maoz,
JHEP {\bf 0212} (2002) 055
[arXiv:hep-th/0012025].

\bibitem 
{M0} S.~D.~Mathur,
arXiv:hep-th/0101118.

\bibitem 
{LM1} O.~Lunin and S.~D.~Mathur,
Nucl.\ Phys.\ B {\bf 610} (2001) 49
[arXiv:hep-th/0105136];\\
Nucl.\ Phys.\ B {\bf 615} (2001) 285
[arXiv:hep-th/0107113];\\
Nucl.\ Phys.\ B {\bf 623} (2002) 342
[arXiv:hep-th/0109154].

\bibitem 
{LM2} O.~Lunin and S.~D.~Mathur,
Nucl.\ Phys.\ B {\bf 642} (2002) 91
[arXiv:hep-th/0206107].

\bibitem 
{LM3} O.~Lunin, S.~D.~Mathur and A.~Saxena,
Nucl.\ Phys.\ B {\bf 655} (2003) 185
[arXiv:hep-th/0211292].

\bibitem 
{LMM} O.~Lunin, J.~Maldacena and L.~Maoz,
arXiv:hep-th/0212210.

\bibitem 
{HE} H.~Elvang,
arXiv:hep-th/0305247.

\bibitem 
{bmpv} J.~C.~Breckenridge, R.~C.~Myers, A.~W.~Peet and C.~Vafa,
Phys.\ Lett.\ B {\bf 391} (1997) 93
[arXiv:hep-th/9602065].

\bibitem
{tseytlin}
A.~A.~Tseytlin,
Mod.\ Phys.\ Lett.\ A {\bf 11}, 689 (1996)
[arXiv:hep-th/9601177].

\bibitem 
{hongteo} K.~Hong and E.~Teo,
Class.\ Quant.\ Grav.\  {\bf 20} (2003) 3269
[arXiv:gr-qc/0305089].

\bibitem 
{rob} R.~C.~Myers,
Phys.\ Rev.\ D {\bf 60} (1999) 046002
[arXiv:hep-th/9903203].

\bibitem 
{blmpsv} J.~C.~Breckenridge, D.~A.~Lowe, R.~C.~Myers, A.~W.~Peet, 
A.~Strominger and C.~Vafa,
Phys.\ Lett.\ B {\bf 381} (1996) 423
[arXiv:hep-th/9603078].

\bibitem 
{cv} M.~Cvetic and D.~Youm,
Nucl.\ Phys.\ B {\bf 476} (1996) 118
[arXiv:hep-th/9603100].

\bibitem 
{HS}
S.~F.~Hassan and A.~Sen,
Nucl.\ Phys.\ B {\bf 375}, 103 (1992)
[arXiv:hep-th/9109038].

\bibitem 
{HHS}
J.~H.~Horne, G.~T.~Horowitz and A.~R.~Steif,
Phys.\ Rev.\ Lett.\  {\bf 68}, 568 (1992)
[arXiv:hep-th/9110065].

\bibitem 
{DNP}
M.~J.~Duff, B.~E.~Nilsson and C.~N.~Pope,
Phys.\ Lett.\ B {\bf 163}, 343 (1985).

\bibitem 
{oldsoln}
R.~Emparan, H.~S.~Reall and E.~Teo, unpublished.

\bibitem 
{misner}
C.~W.~Misner,
J.\ Math.\ Phys.\ {\bf 4}, 924 (1963).

\bibitem 
{ch1} C.~A.~Herdeiro,
Nucl.\ Phys.\ B {\bf 582} (2000) 363
[arXiv:hep-th/0003063].

\bibitem{ME}
E.~J.~Martinec and W.~McElgin,
JHEP {\bf 0210} (2002) 050
[arXiv:hep-th/0206175];\\
JHEP {\bf 0204} (2002) 029
[arXiv:hep-th/0106171].



\bibitem
{emKK} G.~W.~Gibbons and D.~L.~Wiltshire,
Annals Phys.\  {\bf 167} (1986) 201
[Erratum-ibid.\  {\bf 176} (1987) 393].\\
R.~R.~Khuri and T.~Ort{\'\i}n,
Phys.\ Lett.\ B {\bf 373} (1996) 56
[arXiv:hep-th/9512178].

\bibitem
{shein} H.~J.~Sheinblatt,
Phys.\ Rev.\ D {\bf 57} (1998) 2421
[arXiv:hep-th/9705054].

\bibitem 
{GL} R.~Gregory and R.~Laflamme,
Phys.\ Rev.\ Lett.\  {\bf 70} (1993) 2837
[arXiv:hep-th/9301052];
Nucl.\ Phys.\ B {\bf 428} (1994) 399
[arXiv:hep-th/9404071].

\bibitem{DE}
S.~Dimopoulos and R.~Emparan,
Phys.\ Lett.\ B {\bf 526} (2002) 393
[arXiv:hep-ph/0108060].


\bibitem{3m5}
A.~A.~Tseytlin,
Nucl.\ Phys.\ B {\bf 475} (1996) 149
[arXiv:hep-th/9604035].

\bibitem{tub}
R.~Emparan,
Nucl.\ Phys.\ B {\bf 610} (2001) 169
[arXiv:hep-th/0105062].

\end{thebibliography}
\end{document}